\documentclass{LMCS}

\usepackage{hyperref}
\usepackage{enumerate}

\usepackage{ifpdf}
\ifpdf
\usepackage[pdftex,matrix,arrow,tips,curve,color]{xy}
\else
\usepackage[dvips,matrix,arrow,tips,curve,color]{xy}
\fi
\usepackage{xspace}

\usepackage{amssymb}
\usepackage{subfig}

\theoremstyle{plain}
\newtheorem{claim}{Claim}

\theoremstyle{definition}
\newtheorem{notation}[thm]{Notation}

\SelectTips{cm}{10}

\usepackage[english]{babel}

\newcommand{\trsp}[3]{\mathcal{#1} = (#2, #3)}

\newcommand{\rew}{\rightarrow}
\newcommand{\rewt}{\rightarrow^*}
\newcommand{\trewt}{\twoheadrightarrow}
\newcommand{\trewtb}{\twoheadleftarrow}
\newcommand{\trewtp}[1]{\twoheadrightarrow^{#1}}
\newcommand{\out}{\trewt^\mathrm{out}}
\newcommand{\outs}{\rew^\mathrm{out}}

\newcommand{\trewtc}{\mathrel{(\mathord{\trewtb}\mathord{\cdot}\mathord{\trewt})^*}}

\newcommand{\simhc}{\sim_{hc}}
\newcommand{\simhcv}{\wr_{_{\scriptstyle hc}}}

\newcommand{\pos}[1]{\mathcal{P}os(#1)}

\newcommand{\rs}[1]{root(#1)}

\newcommand{\natnum}{\mathbb{N}}

\newcommand{\seper}{\; | \;}

\newcommand{\ulam}{\underline{\lambda}}

\newcommand{\iLC}{i$\lambda$c\xspace}

\newcommand{\dev}{\Rightarrow}

\newcommand{\pmap}{\varepsilon}
\newcommand{\pmapp}[2]{\pmap_{#1}({#2})}

\newcommand{\pme}{\mu}
\newcommand{\pmep}[2]{\pme_{#1}({#2})}

\def\doi{5 (4:3) 2009}
\lmcsheading%
{\doi}
{1--29}
{}
{}
{Oct.~28, 2008}
{Dec.~20, 2009}
{}   

\begin{document}
 
\title[Infinitary Combinatory Reduction Systems: Confluence]{Infinitary Combinatory Reduction Systems:\\
Confluence\rsuper*}

\author[J.~Ketema]{Jeroen Ketema\rsuper a}
\address{{\lsuper a}Research Institute of Electrical Communication, Tohoku University\\
2-1-1 Katahira, Aoba-ku, Sendai 980-8577, Japan}
\email{jketema@nue.riec.tohoku.ac.jp}
\thanks{{\lsuper a}This author was partially funded by the Netherlands Organisation for Scientific Research (NWO) under FOCUS/BRICKS grant number 642.000.502.}

\author[J.~G.~Simonsen]{Jakob Grue Simonsen\rsuper b}
\address{{\lsuper b}Department of Computer Science, University of Copenhagen (DIKU)\\
Universitetsparken 1, 2100 Copenhagen \O, Denmark}
\email{simonsen@diku.dk}

\titlecomment{{\lsuper*}Parts of this paper have previously appeared as \cite{JJ05b}}

\begin{abstract}
We study confluence in the setting of higher-order infinitary rewriting, in particular for infinitary Combinatory Reduction Systems (iCRSs). We prove that 
fully-extended, orthogonal iCRSs are confluent modulo identification of hypercollapsing subterms. As a corollary, we obtain that fully-extended, orthogonal iCRSs have the normal form property and the unique normal form property (with respect to reduction). We also show that, unlike the case in first-order infinitary rewriting, almost non-collapsing iCRSs are not necessarily confluent.
\end{abstract}

\maketitle

\tableofcontents

\section{Introduction}

\noindent This paper is part of a series outlining the fundamental
theory of higher-order infinitary rewriting in the guise of
\emph{infinitary Combinatory Reduction Systems} (iCRSs). In
preliminary papers \cite{JJ05a,JJ05b} we outlined basic motivation and
definitions, and gave a number of introductory results. Moreover, we
lifted a number of results from first-order infinitary rewriting to
the setting of iCRSs. In particular, staple results such as
compression and existence of complete developments of sets of redexes
(subject to certain conditions) were proved.

The purpose of iCRSs is to extend infinitary term rewriting to encompass higher-order rewrite systems.
This allows us, for instance, to reason about the behaviour of the well-known $\mathtt{map}$ functional 
when it is applied to infinite lists. The $\mathtt{map}$ functional
and the usual constructors and destructors for lists can be represented by the below iCRS:
\begin{align*}
\mathtt{map}([z]F(z),\mathtt{cons}(X,XS)) & \rew
\mathtt{cons}(F(X),\mathtt{map}([z]F(z),XS)) \\
\mathtt{map}([z]F(z),\mathtt{nil}) & \rew \mathtt{nil}\\
\mathtt{hd}(\mathtt{cons}(X,XS)) & \rew X \\
\mathtt{tl}(\mathtt{cons}(X,XS)) & \rew XS
\end{align*}

Systems such the above may satisfy certain simple criteria: being \emph{orthogonal} (rules do not overlap syntactically) and \emph{fully-extended} (if a variable is bound, then every meta-variable in its scope must be applied to it). We show that systems satisfying these two criteria are confluent modulo identification of a certain class of `meaningless' subterms: Subterms that are \emph{hypercollapsing}. As an example, $\mathtt{map}$ above, when applied to any infinite list $\mathtt{cons}(s_0,\mathtt{cons}(s_1,\mathtt{cons}(\ldots)))$, will yield identical results no matter how it is computed, except when applied to lists that will never yield a proper result \emph{irrespective} of the evaluation order.

{\bf A succinct description for researchers familiar with infinitary rewriting}: In the current paper, we employ the methods developed in previous papers to show that fully-extended, orthogonal iCRSs are confluent modulo identification of hypercollapsing subterms. As a corollary, we obtain that fully-extended, orthogonal iCRSs have the normal form property and the unique normal form property (with respect to reduction). Finally, we show that, unlike the case in first-order infinitary rewriting, almost non-collapsing iCRSs are not necessarily confluent.  

Parts of this paper have previously appeared as \cite{JJ05b}; the current paper \emph{corrects} the results of that paper and \emph{extends} them: We now allow rules with infinite right-hand sides, not only finite right-hand sides.
The present paper requires some of the results proved in the previously published, peer-reviewed papers \cite{JJ05a,JJ05b}. A much-updated and extended version of these results is available as \cite{paper_i}.

\subsection{Overview and roadmap to confluence}

The contents of the paper are as follows: Section \ref{sec:preliminaries} introduces preliminary notions. Section \ref{sec:pp} on projection pairs recapitulates in an abstract way the fundamental results on \emph{essential rewrite steps}, the primary method used to prove confluence in the higher-order infinitary setting. Section \ref{sec:confluence} provides
proofs of our main results on confluence.
Section \ref{sec:normal_form_properties} considers
the normal form property, the unique normal form property, and the
unique normal form with respect to reduction property. 
Section \ref{sec:conclusion} concludes.

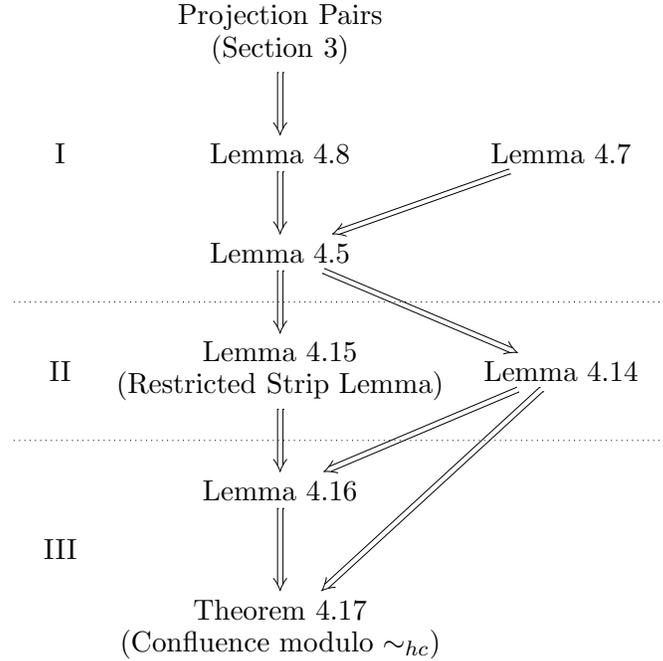
\begin{figure}
\[
\xymatrix@=0.3cm{
  & &  \txt{Projection Pairs\\ (Section  \ref{sec:pp})} \ar@{=>}[dd] &  \\
\\
 & \txt{I} & \txt{Lemma \ref{lem:root_collapses_preserved}} \ar@{=>}[dd] & \txt{Lemma \ref{lem:HCR_is_preserved} } \ar@{=>}[ldd] & \\
\\
 & & \txt{Lemma  \ref{lem:HCR->HC}} \ar@{=>}[dd] \ar@{=>}[ddr] & \\
 \ar@{.}[rrrr] & & & &  \\
 & \txt{II} & \txt{Lemma  \ref{lem:out_strip}\\ (Restricted Strip Lemma)} \ar@{=>}[dd] & \txt{Lemma \ref{hcoutlemma}} \ar@{=>}[ddddl] \ar@{=>}[ddl] && \\
 \ar@{.}[rrrr] & & & & \\
 & & \txt{Lemma \ref{lem:out_implies_closed}} \ar@{=>}[dd] &   \\
 & \txt{III} \\
 & & \txt{Theorem  \ref{the:confl_modulo}\\(Confluence modulo $\simhc$)} & 
}
\]
\caption{\label{fig:road}Roadmap to confluence}
\end{figure}

The main result of the paper is Theorem \ref{the:confl_modulo}:
Fully-extended, orthogonal iCRSs are confluent modulo identification of hypercollapsing subterms. To aid the reader we give a roadmap of the most important auxiliary results leading up to that theorem in Figure \ref{fig:road}.

The auxiliary results are divided into three parts that all depend on the concept of projection pairs (and also the results of Sections \ref{sec:dev} and \ref{sec:tile} concerning developments and tiling diagrams, although not depicted explicitly). Part I, forming Section~\ref{sec:hypercollapsingness}, relates hypercollapsing subterms and so-called hypercollapsing reductions (Lemma  \ref{lem:HCR->HC}). These reductions simplify the reasoning regarding hypercollapsing subterms in the face of the arbitrary reductions that occur in the context of any confluence theorem.

Part II, forming the first half of Section \ref{sec:cm}, considers reductions that do not affect hypercollapsing subterms and establishes a Strip Lemma for such reductions. Although not depicted in Figure \ref{fig:road}, Part II also establishes --- in Proposition \ref{prop:equiv} --- that the relation obtained by replacing the hypercollapsing subterms of a term by other hypercollapsing terms yields an equivalence relation (denoted by $\simhc$).

Part III, forming the latter half of Section \ref{sec:cm}, establishes our confluence theorem. The bulk of the work in this part consist in proving that confluence holds in case reductions that do not affect hypercollapsing subterms are considered (Lemma \ref{lem:out_implies_closed}). Constructing tiling diagrams, the proof heavily depends on the restricted Strip Lemma established in Part II, and thus follows the lines of earlier confluence proofs \cite{T03}. However, the proof also contains a completely novel ingredient: The constructed tiling diagrams are in a sense incomplete and must be superimposed to effectively complete each other. The main result is established in Theorem \ref{the:confl_modulo}.

\section{Preliminaries}
\label{sec:preliminaries}

\noindent We presuppose a working knowledge of the basics of ordinary
finitary term rewriting~\cite{T03}. The basic theory of infinitary
Combinatory Reduction Systems has been laid out in \cite{JJ05a,JJ05b},
and we give only the briefest of definitions in this section. Full
proofs of all results may be found in the above-mentioned
papers. Moreover, the reader familiar with \cite{paper_i} may safely
skip this section; this section is essentially an abstract of that
paper.

Throughout, infinitary Term Rewriting Systems are invariably abbreviated as iTRSs and infinitary $\lambda$-calculus is abbreviated as \iLC. Moreover, we denote the first infinite ordinal by $\omega$, and arbitrary ordinals by $\alpha$, $\beta$, $\gamma$, and so on. We use $\natnum$ to denote the set of natural numbers, starting from zero.

\subsection{Terms, meta-terms, and positions}

We assume a signature $\Sigma$, each element of which has finite arity. We also assume a countably infinite set of variables and, for each finite arity, a countably infinite set of meta-variables of that arity. Countably infinite sets suffice, given that we can employ `Hilbert hotel'-style renaming. 

The (infinite) \emph{meta-terms} are defined informally in a top-down fashion by the following rules, where $s$ and $s_1$, \ldots, $s_n$ are again meta-terms:
\begin{enumerate}[(1)]
\item
each variable $x$ is a meta-term,
\item
if $x$ is a variable and $s$ is a meta-term, then $[x] s$ is a meta-term,
\item
if $Z$ is a meta-variable of arity $n$, then $Z(s_1, \ldots, s_n)$ is a meta-term,
\item
if $f \in \Sigma$ has arity $n$, then $f(s_1, \ldots, s_n)$ is a meta-term.
\end{enumerate}

\noindent We consider meta-terms modulo $\alpha$-equivalence.

A meta-term of the form $[x] s$ is called an
\emph{abstraction}. Each occurrence of the variable $x$ in $s$ is
\emph{bound} in $[x]s$, and each subterm of $s$ is said to occur in the
\emph{scope} of the abstraction. If $s$ is a meta-term, we denote by $\rs{s}$ the root symbol of $s$. Following the definition of meta-terms, we define $\rs{x} = x$, $\rs{[x]s} = [x]$, $\rs{Z(s_1, \ldots, s_n)} = Z$, and $\rs{f(s_1, \ldots, s_n)} = f$.

The set of \emph{terms} is defined as the set of all meta-terms without meta-variables. Moreover, a \emph{context} is defined as a meta-term over $\Sigma \cup \{ \Box \}$ where $\Box$ is a fresh nullary function symbol and a \emph{one-hole} context is a context in which precisely one $\Box$ occurs. If $C[\Box]$ is a one-hole context and $s$ is a term, we obtain a term by replacing $\Box$ by s; the new term is denoted by $C[s]$.

Replacing a hole in a context does not avoid the capture of free variables: A free variable $x$ in $s$ is bound by an abstraction over $x$ in $C[\Box]$ in case $\Box$ occurs in the scope of the abstraction. This behaviour is not obtained automatically when working modulo $\alpha$-equivalence: It is \emph{always} possible find a representative from the $\alpha$-equivalence class of $C[\Box]$ that does not capture the free variables in $s$. Therefore, we will always work with \emph{fixed} representatives from $\alpha$-equivalence classes of contexts. This convention ensures that variables will be captured properly.

\begin{rem}
Capture avoidance is disallowed for contexts as we do not want to lose variable bindings over rewrite steps in case: (i) an abstraction occurs in a context, and (ii) a variable bound by the abstraction occurs in a subterm being rewritten. Note that this means that the representative employed as the context must already be fixed \emph{before} performing the actual rewrite step.

As motivation, consider $\lambda$-calculus: In the term
$\lambda x . (\lambda y . x) z$, contracting the redex 
inside the context $\lambda x . \Box$ yields
$\lambda x.x$, whence the substitution rules for contexts should
be such that 
\[
(\lambda x. \Box)\{(\lambda y . x) z / \Box\}
\rew_\beta \lambda x.x \, .
\]
 If we assumed capture avoidance
in effect for contexts, we would have an $\alpha$-conversion
in the rewrite step, whence
\[
(\lambda x .\Box)\{(\lambda y . x) z / \Box\}
\rew_\beta \lambda w . x \, ,
\]
which is clearly wrong.
\end{rem}

Formally, meta-terms are defined by taking the metric completion of the set of finite meta-terms, the set inductively defined by the above rules. The distance between two terms is either taken as $0$, if the terms are $\alpha$-equivalent, or as $2^{-k}$ with $k$ the minimal depth at which the terms differ, also taking into account $\alpha$-equivalence. By definition of metric completion, the set of finite meta-terms is a subset of the set of meta-terms. Moreover, the metric on finite meta-terms extends uniquely to a metric on meta-terms.

\begin{exa}
Any finite meta-term, e.g.\ $[x]Z(x,f(x))$, is a meta-term. We also have that $Z'(Z'(Z'(\ldots)))$ is a meta-term, as is $Z_1([x_1]x_1,Z_2([x_2]x_2,\ldots))$.

The meta-terms $[x]Z(x,f(x))$ and $[y]Z(y,f(y))$ have distance $0$ and the meta-terms $[x]Z(x,f(x))$ and $[y]Z(y,f(z))$ have distance $\frac{1}{8}$.
\end{exa}

Positions of meta-terms are defined by considering such terms in a top-down fashion. Given a meta-term $s$, its \emph{set of positions}, denoted $\pos{s}$, is the set of finite strings over $\natnum$, with $\epsilon$ the empty string, such that:
\begin{enumerate}[(1)]
\item
if $s = x$ for some variable $x$, then $\pos{s} = \{ \epsilon \}$,
\item
if $s = [x] t$, then $\pos{s} = \{ \epsilon \} \cup \{ 0 \cdot p \seper p \in \pos{t} \}$,
\item
if $s = Z(t_1, \ldots, t_n)$, then $\pos{s} = \{ \epsilon \} \cup \{ i \cdot p \seper 1 \leq i \leq n, \, p \in \pos{t_i} \}$,
\item
if $s = f(t_1, \ldots, t_n)$, then $\pos{s} = \{ \epsilon \} \cup \{ i \cdot p \seper 1 \leq i \leq n, \, p \in \pos{t_i} \}$.
\end{enumerate}

The \emph{depth} of a position $p$, denoted $\vert p \vert$, is the
number of characters in $p$. Given $p, \, q \in \pos{s}$, we write $p \leq q$
and say that $p$ is a \emph{prefix} of $q$, if there exists an $r \in \pos{s}$ such that $p \cdot r = q$. If
$r \not = \epsilon$, we also write $p < q$ and say that the prefix is
\emph{strict}. Moreover, if neither $p \leq q$ nor $q \leq p$, we say that $p$
and $q$ are \emph{parallel}, which we write as $p \parallel q$. 

We denote by $s|_p$ the subterm of $s$ that occurs \emph{at} position $p \in \pos{s}$. Moreover, if $q \in \pos{s}$ and $p < q$, we say that the subterm at position $p$ occurs \emph{above} $q$. Finally, if $p > q$, then we say that the subterm occurs \emph{below} $q$.

Below we introduce a restriction on meta-terms called the \emph{finite chains property}, which enforces the proper behaviour of valuations. Intuitively, a \emph{chain} is a sequence of contexts in a meta-term occurring `nested right below each other'.
\begin{defi}
Let $s$ be a meta-term.  A \emph{chain} in $s$ is a sequence
of (context, position)-pairs $(C_i[\Box],p_i)_{i < \alpha}$, with
$\alpha \leq \omega$, such that for each $(C_i[\Box], p_i)$:
\begin{enumerate}[(1)]
\item
if $i + 1 < \alpha$, then $C_i[\Box]$ has \emph{one} hole and $C_i[t_i] = s|_{p_i}$ for some term $t_i$, and
\item
if $i + 1 = \alpha$, then $C_i[\Box]$ has \emph{no} holes and $C_i[\Box] = s|_{p_i}$,
\end{enumerate}
and such that $p_{i+1} = p_i \cdot q_i$ for all $i + 1 < \alpha$ where $q_i$ is the position of the hole in $C_i[\Box]$.

If $\alpha < \omega$, respectively
$\alpha = \omega$, then the chain is called \emph{finite},
respectively \emph{infinite}.
\end{defi}
\noindent Observe that at most one $\Box$ occurs in any context $C_i[\Box]$ in a chain. In fact, $\Box$ only occurs in $C_i[\Box]$ if $i + 1 < \alpha$; if $i + 1 = \alpha$, we have $C_i[\Box] = s|_{p_i}$.

\subsection{Valuations}

We next define valuations, the iCRS analogue of substitutions as defined for iTRSs and \iLC. As it turns out, the most straightforward and liberal definition of meta-terms has rather poor properties: Applying a valuation
need not necessarily yield a well-defined term. Therefore, we also introduce an important restriction on meta-terms: the \emph{finite chains property}. This property will also prove crucial in obtaining positive results later in the paper.

Essentially, the definitions are the same as in the case of CRSs \cite{KOR93,T03_R}, except that the interpretation of the definition is top-down (due to the presence of infinite terms and meta-terms). Below, we use $\vec{x}$ and $\vec{t}$ as short-hands for, respectively, the sequences $x_1, \ldots, x_n$ and $t_1, \ldots, t_n$ with $n \geq 0$. Moreover, we assume $n$ fixed in the next two definitions.
\begin{defi}
\label{defsubstitution}
A \emph{substitution} of terms 
$\vec{t}$ for distinct variables 
$\vec{x}$ in a term $s$, denoted $s[\vec{x} := \vec{t}]$, is defined as:
\begin{enumerate}[(1)]
\item
$x_i[\vec{x} := \vec{t}] = t_i$,
\item
$y[\vec{x} := \vec{t}] = y$, if $y$ does not occur in $\vec{x}$,
\item
$([y]s')[\vec{x} := \vec{t}] = [y](s'[\vec{x} := \vec{t}])$,
\item
$f(s_1, \ldots, s_m)[\vec{x} := \vec{t}] = f(s_1[\vec{x} := \vec{t}],
\ldots, s_m[\vec{x} := \vec{t}])$.
\end{enumerate}
\end{defi}
\noindent The above definition implicitly takes into account the usual variable
convention \cite{B85} in the third clause to avoid the binding of free
variables by the abstraction. We now define substitutes (adopting this
name from Kahrs \cite{K93}) and valuations.

\begin{defi}
\label{defsubstitute}
An \emph{$n$-ary substitute} is a mapping denoted $\ulam x_1, \ldots, x_n . s$ or $\ulam \vec{x} . s$, with $s$ a term, such that:
\begin{equation}
\label{pbetaeq}
(\ulam \vec{x} . s)(t_1, \ldots, t_n) = s[\vec{x} := \vec{t}] \, .
\end{equation}
\end{defi}

The intention of a substitute is to ensure that proper `housekeeping' 
of substitutions is observed when performing a rewrite step. Reading Equation (\ref{pbetaeq}) from left to right yields a rewrite rule:
\[
(\ulam \vec{x}. s)(t_1, \ldots, t_n) \rew s[\vec{x} := \vec{t}] \, .
\]
The rule can be seen as a \emph{parallel $\beta$-rule}. That is, a variant of the $\beta$-rule from (infinitary) $\lambda$-calculus which simultaneously substitutes multiple variables.

\begin{defi}
\label{defvaluation}
Let $\sigma$ be a function that maps meta-variables to substitutes
such that, for all $n \in \mathbb{N}$, if $Z$ has arity $n$, then so
does $\sigma(Z)$.

A \emph{valuation} induced by $\sigma$ is a relation $\bar{\sigma}$ that takes meta-terms to terms such that:
\begin{enumerate}[(1)]
\item
$\bar{\sigma}(x) = x$,
\item
$\bar{\sigma}([x]s) = [x](\bar{\sigma}(s))$,
\item
$\bar{\sigma}(Z(s_1, \ldots, s_m)) = \sigma(Z)(\bar{\sigma}(s_1), \ldots, \bar{\sigma}(s_m))$,
\item
$\bar{\sigma}(f(s_1, \ldots, s_m)) = f(\bar{\sigma}(s_1), \ldots, \bar{\sigma}(s_m))$.
\end{enumerate}

\end{defi}

Similar to Definition \ref{defsubstitution}, the above definition implicitly takes into account the variable convention, this time in the second clause, to avoid the binding of free variables by the abstraction.

The definition of a valuation yields a straightforward two-step way of applying it to a meta-term: In the first step each subterm of the form $Z(t_1, \ldots, t_n)$ is replaced by a subterm of the form $(\ulam \vec{x}. s)(t_1, \ldots, t_n)$. In the second step Equation \eqref{pbetaeq} is applied to each of these subterms.

In the case of (finite) CRSs, valuations are always (everywhere defined) \emph{maps} taking each meta-term to a unique term \cite[Remark II.1.10.1]{K80}. This is no longer the case when infinite meta-terms are considered. For example, given the meta-term $Z(Z(\ldots Z(\ldots)))$ and applying any map that satisfies $Z \mapsto \ulam x.x$, we obtain $(\ulam x. x)((\ulam x. x)(\ldots (\ulam x. x)(\ldots)))$. Viewing Equation \eqref{pbetaeq} as a rewrite rule, this `$\ulam$-term' reduces only to itself and never to a \emph{term}, as required by the definition of valuations (for more details, see \cite{JJ05a}). To mitigate this problem a subset of the set of meta-terms is introduced in \cite{JJ05a}.
\begin{defi}
Let $s$ be a meta-term.  A \emph{chain of meta-variables} in $s$ is a chain in $s$, written $(C_i[\Box],p_i)_{i < \alpha}$ with $\alpha \leq \omega$, such that for each $i < \alpha$ it is the case that $C_i[\Box] = Z(t_1, \ldots, t_n)$ with $t_j = \Box$ for exactly one $1 \leq j \leq n$.

The meta-term $s$ is said to satisfy the \emph{finite chains property} if no infinite chain of meta-variables occurs in $s$.
\end{defi}

\begin{exa}
\label{ex:chain}
The meta-term $[x_1]Z_1([x_2]Z_2(\ldots [x_n]Z_n(\ldots)))$ satisfies the finite chains property. The meta-terms $Z(Z(\ldots Z(\ldots)))$ and $Z_1(Z_2(\ldots Z_n(\ldots)))$ do not.
\end{exa}

From \cite{JJ05a} we now have the following result:
\begin{prop}
\label{prop:metasane}
Let $s$ be a meta-term satisfying the finite chains property and
let $\bar{\sigma}$ a valuation. There is a unique term that is the
result of applying $\bar{\sigma}$ to $s$. \qed
\end{prop}

\subsection{Rewrite rules and reductions}
\label{sec:rules}

Having defined terms and valuations, we move on to define rewrite rules and reductions.

\subsubsection{Rewrite rules}

We give a number of definitions that are direct extensions of the
corresponding definitions from CRS theory.

\begin{defi}
A finite meta-term is a \emph{pattern} if each of its meta-variables has distinct bound variables as its arguments. Moreover, a meta-term is \emph{closed} if all of its variables occur bound.
\end{defi}

We next define rewrite rules and iCRSs. The definitions are identical to the definitions in the finite case, with exception of the restrictions on the right-hand sides of the rewrite rules: The finiteness restriction is lifted and the finite chains property is put in place. 
\begin{defi}
A \emph{rewrite rule} is a pair $(l, r)$, denoted $l \rew r$, where $l$ is a finite meta-term and $r$ is a meta-term, such that:
\begin{enumerate}[(1)]
\item
$l$ is a pattern with a function symbol at the root,
\item
all meta-variables that occur in $r$ also occur in $l$, 
\item
$l$ and $r$ are closed, and
\item
$r$ satisfies the finite chains property.
\end{enumerate}
The meta-terms $l$ and $r$ are called, respectively, the \emph{left-hand side}
and the \emph{right-hand side} of the rewrite rule.

An \emph{infinitary Combinatory Reduction System (iCRS)} is a pair
$\trsp{C}{\Sigma}{R}$ with $\Sigma$ a signature and $R$ a set of rewrite rules.
\end{defi}

With respect to the left-hand sides of rewrite rules, it is always the case that only finite chains of meta-variables occur, as the left-hand sides are finite.

We now define rewrite steps.
\begin{defi}
A \emph{rewrite step} is a pair of terms $(s, t)$, denoted $s \rew t$, adorned with a one-hole context $C[\Box]$, a rewrite rule $l \rew r$, and a valuation $\bar{\sigma}$ such that $s = C[\bar{\sigma}(l)]$ and $t = C[\bar{\sigma}(r)]$.
The term
$\bar{\sigma}(l)$ is called an \emph{$l \rew r$-redex}, or
simply a \emph{redex}. The redex \emph{occurs} at position $p$ and
depth $|p|$ in $s$, where $p$ is the position of the hole in $C[\Box]$.

A position $q$ of $s$ is said to occur in the \emph{redex pattern} of
the redex at position $p$ if $q \geq p$ and if there does
not exist a position $q'$ with $q \geq p \cdot q'$ such that $q'$ is the
position of a meta-variable in $l$.
\end{defi}

For example, $f([x]Z(x),Z') \rightarrow Z(Z')$ is a rewrite rule,
and $f([x]h(x),a)$ rewrites to $h(a)$ by contracting the redex of the rule 
$f([x]Z(x),Z') \rightarrow Z(Z')$ occurring at position $\epsilon$, i.e.\ at the root.

We now mention some standard restrictions on rewrite rules 
that we need later in the paper: 

\begin{defi}
A rewrite rule is \emph{left-linear}, if each meta-variable occurs at
most once in its left-hand side. Moreover, an iCRS is \emph{left-linear}
if all its rewrite rules are.
\end{defi}

\begin{defi}
Let $s$ and $t$ be finite meta-terms that have no meta-variables in common. The meta-term $s$ \emph{overlaps} $t$ if there exists a non-meta-variable position $p \in \pos{s}$ and a valuation $\bar{\sigma}$ such that $\bar{\sigma}(s|_p) = \bar{\sigma}(t)$.

Two rewrite rules \emph{overlap} if their left-hand sides overlap and if the overlap does not occur at the root when two copies of the same rule are considered. An iCRS is \emph{orthogonal} if all its rewrite rules are left-linear and no two (possibly the same) rewrite rules overlap.
\end{defi}

In case the rewrite rules $l_1 \rew r_1$ and $l_2 \rew r_2$ overlap
at position $p$, it follows that $p$ cannot be the position of a
bound variable in $l_1$. If it were, we would obtain for some valuation
$\bar{\sigma}$ and variable $x$ that $\bar{\sigma}(l_1|_p) = x = 
\bar{\sigma}(l_2)$, which would imply that $l_2$ does not have
a function symbol at the root, as required by the definition of
rewrite rules.

Moreover, it is easily seen that if two left-linear rules overlap in an infinite term, there is also a finite term in which they overlap. As left-hand sides are
\emph{finite} meta-terms, we may appeal to standard ways of deeming
CRSs orthogonal by inspection of their rules.
We shall do so informally on several occasions in the
remainder of the paper.

\begin{defi}
A rewrite rule is \emph{collapsing} if the root of its right-hand side is a meta-variable. 
Moreover, a redex and a rewrite step are \emph{collapsing}
if the employed rewrite rule is. A rewrite step is \emph{root-collapsing}
if it is collapsing and occurs at the root of a term.
\end{defi}

\begin{defi}
\label{def:fully_extended}
A pattern is \emph{fully-extended} \cite{HP96,O96}, if, for each of
its meta-variables $Z$ and each abstraction $[x]s$ having an
occurrence of $Z$ in its scope, $x$ is an argument of that occurrence
of $Z$. Moreover, a rewrite rule is \emph{fully-extended} if its
left-hand side is and an iCRS is \emph{fully-extended} if all its
rewrite rules are.
\end{defi}

\begin{exa}
The pattern $f(g([x]Z(x)))$ is fully-extended. Hence, so is the rewrite
rule $f(g([x]Z(x))) \rightarrow h([x]Z(x))$. The pattern $g([x]f(Z(x),Z'))$, with $Z'$ occurring in the scope of the abstraction $[x]$, is not fully-extended as $x$ does not occur as an argument of $Z'$.
\end{exa}

\subsubsection{Transfinite reductions}

We can now define transfinite reductions.
The definition is equivalent to those for iTRSs and \iLC \cite{KKSV95,KKSV97}.
\begin{defi}
A \emph{transfinite reduction} with domain $\alpha > 0$ is a sequence
of terms $(s_\beta)_{\beta < \alpha}$ adorned with a rewrite step
$s_\beta \rew s_{\beta + 1}$ for each $\beta + 1 < \alpha$. In case $\alpha = \alpha' + 1$,
the reduction is \emph{closed} and of length $\alpha'$. In case
$\alpha$ is a limit ordinal, the reduction is called \emph{open} and
of length $\alpha$. The reduction is \emph{weakly continuous} or
\emph{Cauchy continuous} if, for every limit ordinal $\gamma < \alpha$, the distance between $s_\beta$ and $s_\gamma$ tends to $0$ as $\beta$ approaches $\gamma$ from below. The reduction is \emph{weakly convergent} or \emph{Cauchy convergent} if it is weakly continuous and closed.
\end{defi}
\noindent Intuitively, an open transfinite reduction is lacking a well-defined final term, while a closed reduction does have such a term. 

As in \cite{KKSV95,KKSV97,T03_KV}, we prefer to reason about strongly
convergent reductions.

\begin{defi}
Let $(s_\beta)_{\beta < \alpha}$ be a transfinite reduction. For each
rewrite step $s_\beta \rew s_{\beta + 1}$, let $d_\beta$ denote the
depth of the contracted redex. The reduction is \emph{strongly
  continuous} if it is weakly continuous and if, for every limit ordinal $\gamma < \alpha$, the depth $d_\beta$ tends to infinity as $\beta$ approaches $\gamma$ from below. The reduction is \emph{strongly convergent} if strongly continuous and closed.
\end{defi}

\begin{exa}
\label{ex:wconv}
Consider the rewrite rule $f([x]Z(x)) \rew Z(f([x]Z(x)))$ and observe that $f([x]x) \rew f([x]x)$. Define $s_\beta = f([x]x)$ for all $\beta < \omega \cdot 2$. The reduction $(s_\beta)_{\beta < \omega \cdot 2}$, where in each step we contract the redex at the root, is open and weakly continuous. Adding the term $f([x]x)$ to the end of the reduction yields a weakly convergent reduction. Both reductions are of length $\omega \cdot 2$.

The above reduction is not strongly continuous as all contracted redexes occur at the root, i.e.\ at depth $0$. 
In addition, it cannot be extended to a strongly convergent reduction.
However,
the following reduction
\[
f([x]g(x)) \rew g(f([x]g(x)) \rew \cdots \rew g^n(f([x]g(x))) \rew g^{n + 1}(f([x]g(x))) \rew \cdots
\]
is open and strongly continuous. Extending the reduction with the term $g^\omega$, where $g^\omega$ is shorthand for the infinite term $g(g(\ldots g(\ldots)))$, yields a strongly convergent reduction. Both reductions are of length $\omega$.
\end{exa}

\begin{notation}
By $s \trewtp{\alpha} t$, respectively $s \trewtp{\leq \alpha} t$, we
denote a \emph{strongly convergent} reduction of ordinal
length $\alpha$, respectively of ordinal length at most
$\alpha$. By $s \trewt t$ we denote a \emph{strongly convergent}
reduction of arbitrary ordinal length and by $s \rewt t$
we denote a reduction of finite length.
Reductions are usually ranged over by
capital letters such as $D$, $S$, and $T$.
The concatenation of reductions $S$ and $T$ 
is denoted by $S ; T$.
\end{notation}

Note that the concatenation of any finite number of strongly convergent reductions yields a strongly convergent reduction. For strongly convergent reductions, the following is proved in \cite{JJ05a}.
\begin{lem}
\label{depthlem}
If $s \trewt t$, then the number of steps contracting redexes at depths less than $d \in \natnum$ is finite for any $d$ and $s \trewt t$ has countable length. \qed
\end{lem}

The following result \cite{JJ05a} shows that, as in other forms of infinitary rewriting, reductions can always be `compressed' to have length at most $\omega$:

\begin{thm}[Compression]
\label{the:compression}
For every fully-extended, left-linear iCRS, if $s \trewtp{\alpha} t$, then $s \trewtp{\leq \omega} t$. \qed
\end{thm}

\subsubsection{Descendants and residuals}

The twin notions of \emph{descendants} and \emph{residuals} formalise, respectively, ``what happens'' to positions and redexes across reductions. Across a rewrite step, the only positions that can have descendants are those that occur outside the redex pattern of the contracted redex and that are not positions of the variables bound by abstractions in the redex pattern. Across a reduction, the definition of descendants follows from the notion of a descendant across a rewrite step, employing strong convergence in the limit ordinal case. We do not appeal to further details of the definitions in the remainder of this paper and these details are hence omitted. For the full definitions we refer the reader to \cite{JJ05a}.

\begin{notation}
Let $s \trewt t$. Assume $P \subseteq \pos{s}$ and $\mathcal{U}$ a set of redexes in $s$. We denote the descendants of $P$ across $s \trewt t$ by $P/(s \trewt t)$ and the residuals of $\mathcal{U}$ across $s \trewt t$ by $\mathcal{U}/(s \trewt t)$. Moreover, if $P = \{ p \}$ and $\mathcal{U} = \{ u \}$, then we also write $p/(s \trewt t)$ and $u/(s \trewt t)$. Finally, if $s \trewt t$ consists of a single step contracting a redex $u$, then we sometimes write $\mathcal{U}/u$.
\end{notation}

\subsubsection{Reducts}

In addition to descendants and residuals we need a notion of a reduct of a subterm.
\begin{defi}
Let $s_0 \trewt^\alpha s_\alpha$. Moreover, let $p_0 \in \pos{s_0}$ and $p_\alpha \in \pos{s_\alpha}$. The subterm $s_{\alpha}|_{p_\alpha}$ is called a \emph{reduct} of $s_0|_{p_0}$ if for every $\beta \leq \alpha$ there exists a position $q_\beta$ in $s_\beta$ with $q_\alpha = p_\alpha$ such that:
\begin{enumerate}[$\bullet$]
\item
if $\beta = 0$, then $q_0 = p_0$,
\item
if $\beta = \beta' + 1$, then $q_\beta = q_{\beta'}$ unless $s_{\beta'} \rew s_{\beta' + 1}$ contracts a redex strictly above $q_{\beta'}$ in which case $q_\beta \in q_{\beta'} / (s_{\beta'} \rew s_{\beta' + 1})$, and
\item
if $\beta$ is a limit ordinal, then $q_\beta = q_\gamma$ for all large enough $\gamma < \beta$.
\end{enumerate}

A position $q \geq p_\alpha$ in $s_\alpha$ is said to occur in a \emph{reduct} $s_\alpha|_{p_\alpha}$ of $s_0|_{p_0}$ if, for all positions $p_\alpha < p' \leq q$
in $s_\alpha$, the subterm $s_\alpha\vert_{p'}$ is a reduct
of a subterm strictly below $p_0$ in $s_0$.
\end{defi}

The above notion generalises the usual notion of a reduct. The usual notion is obtained by taking the root position for every $q_\beta$. There is a slight difference between reducts and descendants: Contracting a redex at a position $p$ yields a reduct at position $p$, while $p$ does not have a descendant.

Employing the above definition, we obtain the following property with respect to bound variables; a proof can be found in Appendix \ref{app:bound}.

\begin{lem}
\label{lem:funky_bind}
Let $s_0 \trewt^\alpha s_\alpha$ and suppose $u_\alpha$ and $v_\alpha$ in $s_\alpha$ are residuals of redexes in $s_0$. Denote for all $\gamma \leq \alpha$  by $u_\gamma$ and $v_\gamma$, respectively, the unique redexes at positions $p_\gamma$ and $q_\gamma$ in $s_\gamma$ of which $u_\alpha$ and $v_\alpha$ are residuals.
Assume for all $\gamma < \alpha$ that if the step $s_\gamma \rew s_{\gamma + 1}$ contracts a redex at prefix position of $q_\gamma$ then the redex is a residual of a redex in $s_0$. Then, given that a variable bound by an abstraction in the redex pattern of $u_\alpha$ occurs in $v_\alpha$, it follows that (a) $p_0 < q_0$ and (b) $q_\alpha$ occurs in the reduct $s_\alpha|_{p_\alpha}$ of $s_0|_{p_0}$. \qed
\end{lem}

Observe that, as nestings of subterms can only be created by substitution of bound variables, the above lemma precludes nestings from occurring in reducts unless the conditions in the lemma are met.

\subsection{Developments}
\label{sec:dev}

We need some basic facts about developments which we recapitulate now.

Assuming in the remainder of this section that every iCRS is \emph{orthogonal} and that $s$ is a term and $\mathcal{U}$ a set of redexes in $s$, we first define developments:
\begin{defi}
A \emph{development} of $\mathcal{U}$ is a strongly convergent reduction
such that each step contracts a residual of a redex in $\mathcal{U}$. A
development $s  \trewt t$ is called \emph{complete} if $\mathcal{U}/(s \trewt
t) = \emptyset$. Moreover, a development is called $\emph{finite}$ if
$s \trewt t$ is finite.
\end{defi}

A complete development of a set of redexes does not necessarily exist in the infinite case. Consider for example the rule $f(Z) \rew Z$ and the term $f^\omega$. The set of all redexes in $f^\omega$ does not have a complete development: After any (partial) development a residual of a redex in $f^\omega$ always remains at the root of the resulting term. Hence, any complete development will have an infinite number of root-steps and hence is not strongly convergent.

Although complete developments do not always exist, the following results can still be obtained \cite{JJ05b}, where we write $s \dev^\mathcal{U} t$ for the reduction $s \trewt t$ if it is a complete development of the set of redexes $\mathcal{U}$ in $s$.

\begin{lem}
\label{cdacrossd}
If $\mathcal{U}$ has a complete development and if $s \trewt t$ is a (not necessarily complete) development of $\mathcal{U}$, then $\mathcal{U}/(s \trewt t)$ has a complete development. \qed
\end{lem}

\begin{lem}
Let $s$ be a term and $\mathcal{U}$ a set of redexes in $s$. If $\mathcal{U}$ is finite, then it has a finite complete development. \qed
\end{lem}

\begin{prop}
\label{prop:one_more}
Let $\mathcal{U}$ and $\mathcal{V}$ be sets of redexes in $s$
such that $\mathcal{U}$ has a complete development $s \dev t$ and 
$\mathcal{V}$ is finite. The following diagram commutes:
\[
\xymatrix{
s \ar@{=>}[r]^{\mathcal{V}} \ar@{=>}[d]^{\mathcal{U}} & t'
\ar@{=>}[d]^{\mathcal{U}/(s \dev^\mathcal{V} t')} \\
t \ar@{=>}[r]_{\mathcal{V}/(s \dev^\mathcal{U} t)} & s'
}
\]
\end{prop}
We remark that we do not use the full power of the above proposition: In the current paper $\mathcal{V}$ is always a singleton set.

\subsection{Tiling diagrams}
\label{sec:tile}

Tiling diagrams are defined as follows.

\begin{defi}
A \emph{tiling diagram} of two strongly convergent reductions
$S : s_{0, 0} \rew^{\alpha} s_{\alpha, 0}$ and
$T: s_{0, 0} \rew^{\beta} s_{0, \beta}$ is a rectangular arrangement
of strongly convergent reductions as depicted in Figure \ref{fig:tile} such that (1) each reduction $S_{\gamma,\delta} : s_{\gamma,\delta} \trewt
s_{\gamma + 1,\delta}$ is a complete development of a set of redexes of $s_{\gamma,\delta}$, and similarly for $T_{\gamma,\delta} : s_{\gamma,\delta} \trewt s_{\gamma,\delta+1}$, (2) the leftmost vertical reduction is $S$ and the topmost horizontal reduction is $T$, and (3) for each $\gamma$ and $\delta$ the set of redexes developed in $S_{\gamma,\delta}$ is the set of residuals of the redex contracted in $s_{\gamma,0} \rew s_{\gamma + 1, 0}$ across the (strongly convergent) reduction $T_{\gamma, [0,\delta]} : s_{\gamma, 0} \rew s_{\gamma, 1} \rew \cdots s_{\gamma,\delta}$ (symmetrically for $T_{\gamma,\delta}$).

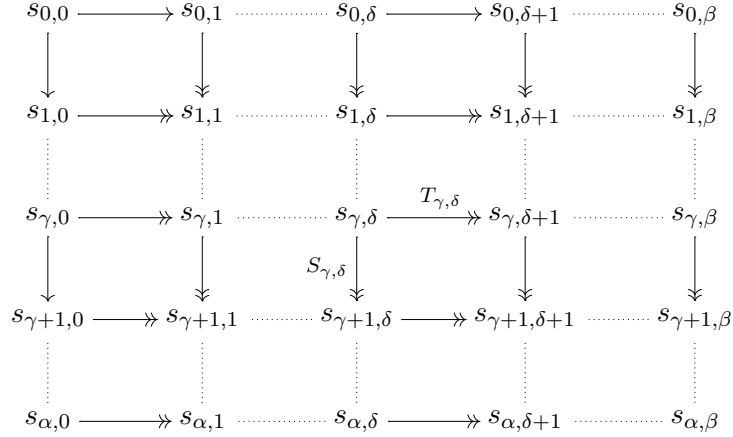
\begin{figure}
\[
\xymatrix{
s_{0, 0} \ar[r] \ar[d]
    & s_{0, 1} \ar@{.}[r] \ar@{->>}[d]
    & s_{0, \delta} \ar[r] \ar@{->>}[d]
    & s_{0, \delta + 1} \ar@{.}[r] \ar@{->>}[d]
    & s_{0, \beta} \ar@{->>}[d] \\
s_{1, 0} \ar@{->>}[r] \ar@{.}[d]
    & s_{1, 1} \ar@{.}[r] \ar@{.}[d]
    & s_{1, \delta} \ar@{->>}[r] \ar@{.}[d]
    & s_{1, \delta + 1} \ar@{.}[r] \ar@{.}[d]
    & s_{1, \beta} \ar@{.}[d] \\
s_{\gamma, 0} \ar@{->>}[r] \ar@{->}[d]
    & s_{\gamma, 1} \ar@{.}[r] \ar@{->>}[d]
    & s_{\gamma, \delta} \ar@{->>}[r] ^{T_{\gamma, \delta}}
                         \ar@{->>}[d]_{S_{\gamma, \delta}}
    & s_{\gamma, \delta + 1} \ar@{.}[r] \ar@{->>}[d]
    & s_{\gamma, \beta} \ar@{->>}[d] \\
s_{\gamma + 1, 0} \ar@{->>}[r] \ar@{.}[d]
    & s_{\gamma + 1, 1} \ar@{.}[r] \ar@{.}[d]
    & s_{\gamma + 1, \delta} \ar@{->>}[r] \ar@{.}[d]
    & s_{\gamma + 1,\delta + 1} \ar@{.}[r] \ar@{.}[d]
    & s_{\gamma + 1, \beta} \ar@{.}[d] \\
s_{\alpha, 0} \ar@{->>}[r]
    & s_{\alpha, 1} \ar@{.}[r]
    & s_{\alpha, \delta} \ar@{->>}[r]
    & s_{\alpha, \delta + 1} \ar@{.}[r]
    & s_{\alpha, \beta}
}
\]
\caption{\label{fig:tile}A tiling diagram}
\end{figure}
\end{defi}

For $S_{[0, \alpha], \beta}$ we usually write $S/T$ and we call this reduction the \emph{projection} of $S$ across $T$ (similarly for $T_{\alpha, [0, \beta]}$ and $T/S$). Moreover, if $T$ consists of a single step contracting a redex $u$, we also write $S/u$ (symmetrically $T/u$).

Given two strongly convergent reductions, even in the case these where one of these is \emph{finite}, a tiling need not exist, witness e.g.\ the failure of the Strip Lemma in \cite{KKSV97}. To cope with this issue later in the paper we employ the following theorem from \cite{JJ05b} in combination with the results from Section \ref{sec:dev}. The theorem, which is valid for \emph{orthogonal} iCRSs, extends Theorem 12.6.5 from \cite{T03_KV}: In \cite{T03_KV} it is assumed that $S$ and $T$ are reductions of limit ordinal length; in this paper,
$S$ and $T$ may be reductions of arbitrary ordinal length.

\begin{thm}
\label{the:tiling_completion_convergence}
Let $S$ and $T$ be strongly convergent reductions starting from the same term. Suppose that the tiling diagram for $S$ and $T$ exists except that it is unknown if $S/T$ and $T/S$ are strongly convergent and end in the same term. The following are equivalent:
\begin{enumerate}[\em(1)]
\item
The tiling diagram of $S$ and $T$ can be completed, i.e.\ $S/T$ and $T/S$ are strongly convergent and end in the same term.
\item
$S/T$ is strongly convergent.
\item
$T/S$ is strongly convergent. \qed
\end{enumerate}
\end{thm}

\section{Projection pairs}
\label{sec:pp}

\noindent For the confluence result, we shall employ a technique by
van Oostrom \cite{O99}, combining the concept of essentiality from
\cite{K88,GK94} with a termination technique from \cite{SR90,M97}.  We
give an abstract formulation of the technique in terms of so-called
\emph{projection pairs}; the formulation is taken from \cite{K08} and
extends the more primitive notions from \cite{paper_iii}. Please note
that the main definitions given below do not occur in
\cite{paper_iii}, and the reader is thus advised to review them
carefully.

We require an auxiliary definition:

\begin{defi}
Let $s$ and $t$ be terms and $P \subseteq \pos{s}$. The set $P$ is a \emph{prefix set} of $s$ if $P$ is \emph{finite} and if all prefixes of positions in $P$ are also in $P$. Moreover, $t$ \emph{mirrors} $s$ in $P$, if for all $p \in P$ it holds that $p \in \pos{t}$ and $\rs{t|_p} = \rs{s|_p}$ (modulo $\alpha$-equivalence).
\end{defi}

Van Oostrom's technique uses a termination argument on a prefix set $P$
and a reduction $D$ that consists of a finite sequence of complete developments starting
from a term $s$. The crux of the termination argument is, as always, some measure $\pme$ over a well-founded order that 
decreases across the sequence of developments. 

The technique hinges on projecting $D$ across a single rewrite step starting from $s$. If
the rewrite step occurs in some specific prefix, $Q$, of $s$, it is called \emph{essential};
otherwise it is called \emph{inessential}. Projecting $D$ across the step to obtain a new sequence $D'$, one shows by case analysis that the measure is always non-increasing, but decreases strictly if the step is essential.
The specific prefix, $Q$, is obtained from a prefix set $P$ of the final term of $D$
by a map $\pmap$ mapping $P$ to $Q$. The pair $(\pme,\pmap)$ 
is called a \emph{projection pair}.

Intuition done, we now proceed to give precise definitions:

\begin{defi}
\label{def:pp}
Given a well-founded order $\prec$ on a set $O$, a \emph{projection pair} is a pair $(\pme, \pmap)$ of maps over finite sequences of complete developments $D$ and prefix sets $P$ of the final term of the chosen $D$ such that:
\begin{enumerate}[$\bullet$]
\item
$\pmep{P}{D}$ maps to an element of $O$, and
\item
$\pmapp{P}{D}$ maps to a prefix set of the initial term of $D$,
\end{enumerate}
and such that if $D'$ is a sequence of complete developments strictly shorter than $D$ with $P'$ a prefix set of the final term of $D'$, then $\pmep{P'}{D'} \prec \pmep{P}{D}$.
\end{defi}
\noindent The map $\pme$ is the measure and $\pmap$ is the map for prefix sets. The measure requires a sequence that is strictly shorter than $D$ to map to a smaller element in the well-founded order. Although of a technical nature, this property is easily obtained in case tuples are used to define the well-founded order and the tuples are first compared length-wise and next lexicographically.

We can now define (in)essentiality as follows:
\begin{defi}
Let $(\pme, \pmap)$ be a projection pair. If $D$ is a finite sequence of complete developments and $P$ is a prefix set of the final term of $D$, then a position $p$ of, respectively a redex $u$ in, the initial term of $D$ is called \emph{essential} for $P$ if $p$, respectively the position of $u$, occurs in $\pmapp{P}{D}$. A position, respectively a redex, is called \emph{inessential} otherwise.
\end{defi}

The existence of the projection mentioned above can now be formulated as the soundness of a projection pair:
\begin{defi}
\label{def:ppsound}
Let $\prec$ be a well-founded order on a set $O$. A projection pair $(\pme, \pmap)$ is \emph{sound} if for every finite sequence of complete development $D$, prefix set $P$ of the final term of $D$, and $s \trewt t$, with $s$ the initial term of $D$, it holds that:
\begin{enumerate}[(1)]
\item
if $s \trewt t$ consists of a \emph{single} step contracting a redex $u$ at an essential position, with no residual in $u/D$ occurring at a position in $P$, then there exists a $D'$ such that $\pmep{P}{D'} \prec \pmep{P}{D}$, and
\item
if $s \trewt t$ consists of one or more steps and \emph{only} contracts redexes at \emph{in}essential positions, then there exists a $D'$ such that $\pmep{P}{D'} = \pmep{P}{D}$ and $\pmapp{P}{D'} = \pmapp{P}{D}$,
\end{enumerate}
where in both cases $D'$ is a finite sequence of complete developments with initial term $t$ such that the final term of $D'$ mirrors the final one of $D$ in $P$.
\end{defi}
\noindent The restriction in the first clause that no residual in $u/D$ occurs in $P$ ensures that the projection preserves $P$. Together, the clauses formalise the intuition behind $\pmap$, i.e.\ that $P$ only depends on positions in $\pmapp{P}{D}$. The map is constant for reductions contracting only redexes outside $\pmapp{P}{D}$ and, obviously, any term in such a reduction mirrors all the other terms in $\pmapp{P}{D}$.

\begin{rem}
The first clause of Definition \ref{def:ppsound} deals neither with reductions where residuals from $u/D$ occur in $P$ nor with infinite reductions. In the next section, we deal with the first by means of the restriction on strictly shorter sequences of complete developments and with the second by means of strong convergence.
\end{rem}

We have the following theorem, proved in \cite{paper_iii}:

\begin{thm}
\label{thm:pp}
For each fully-extended, orthogonal iCRS a sound projection pair exists.
\end{thm}

\section{Confluence}
\label{sec:confluence}

\noindent We will now present our confluence result. To start, recall
that confluence in general does not hold for iTRSs, even under
assumption of orthogonality \cite{KKSV95}. As every iTRS can be seen
as a fully-extended iCRS, it follows that fully-extended, orthogonal
iCRSs are in general not confluent either.

In case of iTRSs two approaches are known for restoring confluence \cite{KKSV95}, namely (1) identifying all subterms that disrupt confluence, and (2) restricting the rewrite rules that are allowed. Identifying all subterms that disrupt confluence leads to the definition of so-called hypercollapsing subterms and yields the result that orthogonal iTRSs are confluent modulo these subterms. Restricting the rules that are allowed yields results regarding almost non-collapsing iTRSs.

Considering only  \emph{fully-extended}, \emph{orthogonal} iCRSs, we next prove that such iCRSs are also confluent modulo hypercollapsing subterms, where a term $s$ is called hypercollapsing if for every $s \trewt t$ we have that $t \trewt t'$ where $t'$ has a collapsing redex at the root. This not only generalises the result for iTRSs but also a similar result for \iLC \cite{KKSV97}. Regrettably, the proofs for iTRSs and \iLC from \cite{T03_KV} cannot be lifted to the general higher-order case: For iTRSs the proof hinges on the Strip Lemma and for \iLC it hinges on the notion of head reduction, both of which fail to properly generalise to iCRSs. To circumvent these problems, we employ the measure defined in the previous section.

As an added benefit, we are able to overcome a small infelicity in the similar proof for \iLC in \cite{T03_KV}. There, Lemma 12.8.14 treats reductions 
outside hypercollapsing subterms in a way similar to our Lemma \ref{lem:out_implies_closed}; however, for \iLC, the induction step in the proof of
\cite{T03_KV} can apparently only be carried out if a stronger induction hypothesis is assumed than the one given --- the two resulting reductions should be outside hypercollapsing subterms. The general result for iCRSs 
given in the present paper subsumes the one for \iLC.

Apart from confluence modulo, we show in Section \ref{sec:almost_non_collapsing} that the positive result that an iTRS is confluent if{f} it is almost non-collapsing cannot be trivially lifted to iCRSs.

\begin{rem}
On a historical note: Courcelle \cite{C83} observed similar problems with confluence while trying to define second-order substitutions on infinite trees. He circumvented these problems by requiring rules to be non-collapsing. In a general setting such as ours this would be too harsh a restriction.
\end{rem}

\subsection{Hypercollapsingness}
\label{sec:hypercollapsingness}

We now proceed to define a particularly troublesome kind of reduction and term.

\begin{defi}
\label{def:hyper_red}
A \emph{hypercollapsing reduction} is an open strongly continuous reduction with an infinite number of root-collapsing steps.
\end{defi}
\noindent
Thus, a hypercollapsing reduction is a particular example of a transfinite reduction of some limit ordinal length $\alpha$ that cannot be extended to a strongly \emph{convergent} 
reduction --- the term $s_\alpha$ is undefined. Note that, writing $(s_\beta)_{\beta < \alpha}$ for a hypercollapsing reduction sequence, we have that every initial sequence $(s_\beta)_{\beta < \gamma + 1}$ with $\gamma < \alpha$ \emph{is} strongly convergent.

\begin{exa}
Hypercollapsing reductions are known even in the first-order case
where we have, e.g.\ (in the syntax of iCRSs) the rewrite rule
$f(Z) \rew Z$ and the term $f^\omega$ from which there is the
hypercollapsing reduction
\[
f^\omega \rew f^\omega \rew \cdots
\]
which is obtained by repeatedly contracting the redex at the root.

For an example in more higher-order spirit, consider the rule
$g([x]Z(x)) \rew Z([x]Z(x))$. From the term
$g([x]g(x))$ there is the hypercollapsing
reduction
\[
g([x]g(x)) \rew g([x]g(x)) \rew \cdots \, .
\]
which is again obtained by repeatedly contracting the redex at the root.
\end{exa}

The crucial definition is now the following:

\begin{defi}
A term $s$ is said to be \emph{hypercollapsing} if,
for all terms $t$ with $s \trewt t$, there exists a term
$t'$ with $t \trewt t'$ such that $t'$ has a collapsing redex at the root.
\end{defi}

It is not hard to see that a hypercollapsing term has a
hypercollapsing reduction starting from it; the converse, however, is much
more difficult, and is contained in the following lemma, to the proof of
which we devote the remainder of the section.

\begin{lem}
\label{lem:HCR->HC}
Let $s$ be a term. If there is a hypercollapsing reduction starting from $s$, 
then $s$ is hypercollapsing.
\end{lem}

To start, we observe that 
hypercollapsing reductions satisfy a `compression' property:
\begin{lem}
\label{lem:hycomp}
Let $s$ be a term. If there is a hypercollapsing reduction 
starting from $s$, then there is a hypercollapsing reduction of length
$\omega$ starting from it.
\end{lem}

\proof
By definition, we may write a hypercollapsing reduction starting from $s$ as:
\[
s = s_0 \trewt s'_0 \rew s_1 \trewt s'_1 \rew s_2 \trewt \cdots  \, ,
\]
where $s'_i \rew s_{i + 1}$ is
root-collapsing and no root-collapsing steps occur in
$s_i \trewt s'_i$ for all $i \in \natnum$.

We inductively define a hypercollapsing reduction of length $\omega$:
\[
s = t_0 \rewt t'_0 \rew t_1 \rewt t'_1 \rew t_2 \rewt \cdots  \, ,
\]
where for all $i \in \natnum$ it holds that $t'_i \rew t_{i + 1}$ is root-collapsing and that $t_i \rewt t'_i$ is finite and without root-collapsing steps. First, define $t_0 = s_0 = s$. Next, assume we have defined a term $t_i$ with $t_i \trewt s_i$. Compression of $t_i \trewt s_i \trewt s'_i \rew s_{i + 1}$ yields a reduction $t_i \rewt t'_i \rew t_{i + 1} \trewtp{\leq \omega} s_{i + 1}$ with $t'_i \rew t_{i + 1}$ root-collapsing and $t_i \rewt t'_i$ finite and without root-collapsing steps. Thus, there is a hypercollapsing reduction with the required properties. \qed

The following is the iCRS analogue of Lemma 12.8.4 in \cite{T03_KV} for
iTRSs and strengthening for \iLC:
\begin{lem}
\label{lem:HCR_is_preserved}
Let $s$ and $t$ be terms with $s \rew t$.
If there is a hypercollapsing reduction starting in $s$, then there is a hypercollapsing reduction starting in $t$.
\end{lem}

\proof
Define $s_0 = s$, $t_0 = t$, and suppose $u$ is the redex contracted in $s \rew t$. By Lemma~\ref{lem:hycomp}, we may write the hypercollapsing reduction starting in $s_0$ as:
\[
s_0 \rewt s'_0 \rew s_1 \rewt s'_1 \rew s_2 \rewt \cdots  \, ,
\]
where for all $i \in \natnum$, we have that $s'_i \rew s_{i + 1}$ is root-collapsing and $s_i \rewt s'_i$ is finite and without root-collapsing steps.
By repeated application of Proposition
\ref{prop:one_more}, we obtain the following diagram:
\[
\xymatrix{
s_0 \ar[d]^{u} \ar[r]^{*} & s'_0 \ar@{=>}[d]^{\mathcal{U}'_0} \ar[r] & s_1
\ar@{=>}[d]^{\mathcal{U}_1} \ar[r]^{*} & s'_1 \ar@{=>}[d]^{\mathcal{U}'_1}
\ar[r] & s_2 \ar@{=>}[d]^{\mathcal{U}_2} \ar[r]^{*} & \cdot \ar@{:>}[d]
\ar@{.}[r] & \\
t_0 \ar@{->>}[r] & t'_0 \ar@{->>}[r] & t_1 \ar@{->>}[r] & t'_1 \ar@{->>}[r]
& t_2 \ar@{->>}[r] & \cdot \ar@{.}[r] &
}
\]
Write $S_i$ for $s_i \rewt s'_i \rew s_{i + 1} \rewt \cdots$ and
$T_i$ for $t_i \trewt t'_i \trewt t_{i + 1} \trewt \cdots$.
If we can show for each $i \in
\natnum$ that a root-collapsing step occurs in $T_i$, then an infinite
number of root-collapsing steps occurs in $T_0$, implying that the
reduction is hypercollapsing.

To show that a root-collapsing step occurs in each $T_i$ we distinguish
two cases: (1) a root-collapsing step occurs in $S_i$ not
contracting a residual of $u$, and (2) all root-collapsing steps in $S_i$
contract a residual of $u$. We treat each of these cases in turn:
\begin{enumerate}[(1)]
\item
Suppose a root-step occurs in $S_i$ that does not contract a residual of $u$. Thus, there exists a root-collapsing step $s'_j \rew s_{j + 1}$ with $j \geq i$ such that the contracted redex, say $v$, is not a residual of $u$. Since we have by construction that $\mathcal{U}'_j$ contracts only residuals of $u$, orthogonality implies that a residual of $v$ occurs at the root of $t'_j$ and that no other residuals of $v$ occur in $t'_j$. Also by construction, $t'_j \trewt t_{j + 1}$ contracts precisely all residuals of $v$. Hence, $t'_j \trewt t_{j + 1}$ is a root-collapsing step.
\item
Suppose all root-collapsing steps in $S_i$ contract a residual of $u$ (which implies $u$ is a collapsing redex). Moreover, for any term in $S_i$ call a set $\mathcal{V}$ of residuals of $u$ a \emph{root-nesting} if $\mathcal{V}$ is the largest set such that for each redex $v$ in $\mathcal{V}$ there exists a (partial) development of $\mathcal{V}$ that ends in a term with a residual of $v$ at the root (this residual is also a residual of $u$).

For every term along $S_i$ the root-nesting is finite and non-empty. Finiteness follows as only a finitely many steps occur before each term in $S_i$ and as right-hand sides of rewrite rules satisfy the finite chains condition. Non-emptiness follows as otherwise a root-step occurs that (a) does not contract a residual of $u$ and (b) brings a residual of $u$ to the root. Such a step is by definition root-collapsing, contradicting the assumption that all root-collapsing steps in $S_i$ contract residuals of $u$.

We make the following claim:
\begin{claim}
The number of redexes in a root-nesting eventually increases due to contraction of a step outside the root-nesting.
\end{claim}

To prove the claim, observe that, by definition,
any redex inside a root-nesting occurring at a non-root position occurs
as an argument of another redex inside the root-nesting. As no redex outside the root-nesting occurs above the root-nesting, the cardinality of a root-nesting can, hence, only decrease by contracting a redex inside the root-nesting.

Now suppose the cardinality of the root-nesting increases only by contracting redexes inside the root-nesting itself. By definition of root-nestings, an increase in cardinality is due in this case to nestings that are created among the redexes already present in the root-nesting. By Lemma \ref{lem:funky_bind} and the fact that only a finite number of redexes occur above each other redex, only a finite number of nestings occur that increase the cardinality. Hence, as an infinite number of root-collapsing steps occurs in $S_i$, all of which are in the root-nesting, eventually only decreases can occur, whence, by finiteness of root-nestings, all redexes in the root-nesting must be contracted, contradicting the non-emptiness of root-nestings. This concludes the proof of Claim 1.

By Claim 1, a step outside a root-nesting of $S_i$ occurs that increases the cardinality. The redex contracted in the step, say $v$, is collapsing and does not contract a residual of $u$, by definition of root-nestings. Moreover, as the cardinality increases, a (partial) development of residuals of $u$ exists which brings a residual $v$ to the root. As $v$ is not a residual of $u$, it follows by Lemma \ref{cdacrossd} and the fact that complete developments of residuals of $u$ in terms along $S_i$ exist, that a root-collapsing redex occurs in $T_i$. Since the redex is a residual of a collapsing redex in $S_i$ which is eventually contracted, a root-collapsing step occurs in $T_i$.
\end{enumerate}

As required, we have that a root-step occurs in each $T_i$. Hence, $T_0$ is a hypercollapsing reduction starting from $t_0 = t$. \qed

The next lemma shows that the property of being reducible to a term
with a collapsing redex at the root cannot be destroyed by reductions
unless they contain a collapsing step at the root themselves. In the
proof of the lemma we assume that we have at our disposal a sound projection pair,
as is possible by Theorem \ref{thm:pp}.

\begin{lem}
\label{lem:root_collapses_preserved}
If $s \trewt t$ has no root-collapsing steps and $s$ reduces to a collapsing redex, then so does $t$.
\end{lem}

\proof
We show by ordinal induction that every term $s_\alpha$ in $s \trewt t$ reduces to a collapsing redex by a finite sequence of complete developments $D_\alpha$. Denote by $P_\alpha$ the set of positions of the redex pattern at the root of the final term of $D_\alpha$ and remark that this set is a prefix set. To facilitate the induction we also show for each $\beta \leq \alpha$ either that $\pmep{P_\alpha}{D_\alpha} \prec \pmep{P_\beta}{D_\beta}$ or that $\pmep{P_\alpha}{D_\alpha} = \pmep{P_\beta}{D_\beta}$, $\pmapp{P_\alpha}{D_\alpha} = \pmapp{P_\beta}{D_\beta}$, and $s_\beta \trewt s_\alpha$ consists solely of inessential steps.

For $s_0 = s$, it follows by assumption that $s_0$ reduces to a collapsing redex. In fact, by strong convergence and compression, $s_0$ reduces to a collapsing redex by a finite reduction $D_0$. As any finite reduction can be seen as a finite sequence of complete developments the result follows.

For $s_{\alpha + 1}$, there are two cases to consider given the redex $u$ contracted in $s_\alpha \rew s_{\alpha + 1}$ depending on the occurrence of a residual of $u$ at the root of the final term of $D_\alpha$:
\begin{enumerate}[$\bullet$]
\item
In case no residual of $u$ occurs at the root of the final term of
$D_\alpha$, we discriminate between $u$ being either essential or
inessential for $P_\alpha$. In case $u$ is essential, the result
follows by the induction hypothesis and Definition
\ref{def:ppsound}(1).  Otherwise, the induction hypothesis and
Definition \ref{def:ppsound}(2) can be applied, where the assumed
reduction consists of a single step.

\item
In case a residual of $u$ occurs at the root of the final term of $D_\alpha$, a root-collapsing step not contracting a residual of $u$ occurs somewhere along $D_\alpha$. Otherwise, a residual of $u$ cannot occur at the root of the final term of $D_\alpha$, because $s_\alpha \rew s_{\alpha + 1}$ is not root-collapsing. Hence, there exists a finite sequence $D'_\alpha$ of complete developments that is shorter than $D_\alpha$ and that has a collapsing redex, other than a residual of $u$, at the root of its final term. By Definition \ref{def:pp}, it follows that $\pmep{P'_\alpha}{D'_\alpha} \prec \pmep{P_\alpha}{D_\alpha}$, where $P'_\alpha$ is the set of positions of the redex pattern at the root of the final term of $D'_\alpha$. The case in which no residual of $u$ occurs at the root of the final term of the complete development now applies and the result follows.
\end{enumerate}

For $s_\alpha$ with $\alpha$ a limit ordinal, it follows by the well-foundedness of $\prec$ and the induction hypothesis that there exist a $\beta < \alpha$ such that for every $\beta < \gamma < \alpha$ we have $\pmep{P_\gamma}{D_\gamma} = \pmep{P_\beta}{D_\beta}$. Hence, since we also have by the induction hypothesis that $\pmapp{P_\gamma}{D_\gamma} = \pmapp{P_\beta}{D_\beta}$ for all $\beta < \gamma < \alpha$ and that all redexes contracted in $s_\beta \trewt s_\gamma$ are inessential, the result follows by strong convergence and Definition \ref{def:ppsound}(2), where the assumed reduction is $s_\beta \trewt s_\alpha$. \qed

We can now prove Lemma \ref{lem:HCR->HC}:

\proof[Proof of Lemma \ref{lem:HCR->HC}]
Let $s \trewt t$ be arbitrary. By compression and strong convergence,
we may write $s \rewt t' \trewtp{\leq \omega} t$ such that all
root-reductions occur in $s \rewt t'$. By repeated application of Lemma
\ref{lem:HCR_is_preserved}, there exists a hypercollapsing reduction
starting from $t'$. In particular, $t'$ reduces to a collapsing redex.
Since $t' \trewt t$ contains no steps at the root, Lemma
\ref{lem:root_collapses_preserved} yields that $t$ reduces to a
collapsing redex, proving that $s$ is hypercollapsing. \qed

\subsection{Confluence modulo}
\label{sec:cm}

We now prove confluence modulo identification of hypercollapsing
subterms. Confluence modulo is defined as follows:
\begin{defi}
\label{def:confmod}
An iCRS is \emph{confluent modulo} an equivalence relation
$\sim$ if for all $s \trewt s'$ and $t \trewt t'$ with $s \sim t$ there
exist terms $s''$ and $t''$ such that $s' \trewt s''$ and
$t' \trewt t''$ with $s'' \sim t''$ (see Figure \ref{fig:confmod}).
\end{defi}

\begin{figure}
\[
\xymatrix@=0.5cm{
 & s \ar@{->>}[dl] \ar@{}[r]|*\txt{$\sim$} & t \ar@{->>}[dr] &    \\
s' \ar@{->>}[dr] & & & t' \ar@{->>}[dl] \\
 & s'' \ar@{}[r]|*\txt{$\sim$} & t'' & \\
}
\]
\caption{\label{fig:confmod}Definition \ref{def:confmod}}
\end{figure}

We first show that identification of hypercollapsing subterms yields an equivalence relation. To this end we introduce some notation and show that hypercollapsingness is preserved under replacement of hypercollapsing subterms.
\begin{notation}
We write $s \simhc t$ if $t$ can be obtained from $s$ by replacing a number of hypercollapsing subterms of $s$ by other hypercollapsing terms.
\end{notation}

\begin{prop}
\label{prop:subreplaceokay}
Let $s$ and $t$ be terms. If $s$ is hypercollapsing and $s \simhc t$, then $t$ is hypercollapsing.
\end{prop}

\proof
Let $P$ be the set of positions of hypercollapsing subterms in $s$ that are replaced to obtain $t$. By definition of $s$ there exists a hypercollapsing reduction $S$ starting from it. The redex patterns employed in the steps of $S$ either occur completely outside or completely inside the reducts of subterms in $s$ at positions in $P$. This follows by orthogonality and the fact that the subterms at positions in $P$ are hypercollapsing, i.e.\ each reduct reduces to a term with a collapsing redex at the root. By orthogonality and by the fact that free variables cannot become bound when substituted into the reducts, it does not matter whether any substitutes occur in the reducts.

Omit from $S$ all steps that occur inside the reducts of subterms in $s$ at positions in $P$ to obtain a reduction $S'$ of length $\alpha$. By definition of $S'$, together with orthogonality and fully-extendedness, there exists a reduction $T$ of length $\alpha$ starting in $t$ such that for all $\beta \leq \alpha$ we have that the redex pattern and position of the redex contracted in the $\beta$th step of both $S'$ and $T$ are identical. Hence, if $S'$ is hypercollapsing, then so is $T$ and the result follows by Lemma \ref{lem:HCR->HC}. If $S'$ is not hypercollapsing, then $s$ reduces to a reduct of subterm at a position $p \in P$ and the same holds for $T$. As the subterm at position $p$ in $t$ is hypercollapsing, there exist a hypercollapsing reduction starting from it. Again, by the fact that free variables cannot get bound when terms are substituted into other terms and by orthogonality, it is irrelevant that any substitutes occur. Hence, $T$ can be prolonged to obtain a hypercollapsing reduction and the result follows again by Lemma \ref{lem:HCR->HC}. \qed

We can now prove that $\simhc$ has the required properties:
\begin{prop}
\label{prop:equiv}
The relation $\simhc$ is an equivalence relation, which is closed under substitution of terms for free variables.
\end{prop}

\proof
We have to prove that the relation is reflexive, symmetric, and transitive. Reflexivity and symmetry are immediate by definition. Transitivity follows by Proposition \ref{prop:subreplaceokay}.

To see that relation is closed under substitution, consider a hypercollapsing term $s$ and a term $t$ that is a substitution instance of $s$. By definition of $s$ there exists a hypercollapsing reduction $S$ of length $\alpha$ starting from it. By orthogonality and the fact that no free variables are bound in the terms substituted into $s$, there exists a reduction $T$ of length $\alpha$ starting from $t$ such that for all $\beta \leq \alpha$ we have that the redex pattern and position of the redex contracted in $\beta$th step of both $S$ and $T$ are identical. Hence, since $S$ is hypercollapsing, so is $T$ and the result follows by Lemma \ref{lem:HCR->HC}. \qed

Introducing some further notation, we next show that we can accurately `simulate' reductions in terms that are $\simhc$-related. 

\begin{notation}
By $s \outs t$ we denote a rewrite step that does not occur inside any hypercollapsing subterm of $s$. 
\end{notation}

\begin{lem}
\label{hcoutlemma}
Let $s \trewt t$ have $\alpha$ steps that occur outside hypercollapsing subterms. If $s \simhc s'$, then there exists a reduction $s' \out t'$ of length $\alpha$ such that $t \simhc t'$. Moreover, for all $\beta \leq \alpha$ the redex pattern and position of the redex contracted in the $\beta$th step of  $s' \out t'$ are identical to those of the $\beta$th step of $s \trewt t$ that occurs outside a hypercollapsing subterm.
\end{lem}

\proof
Let $s \trewtp{\gamma} t$ and $s \simhc s'$. We prove the result by ordinal induction on $\gamma$.

If $\gamma = 0$, the result is immediate, as an empty reduction is by definition one that only contracts redexes outside hypercollapsing subterms.

If $\gamma = \delta + 1$, assume $s \trewtp{\gamma} t = s \trewtp{\delta} s_\delta \rew t$. By the induction hypothesis there exists a term $s'_\delta$ such that $s' \trewtp{out} s'_\delta$ and $s_\delta \simhc s'_\delta$. There are two possibilities for $s_\delta \rew t$, depending on the contracted redex occurring either outside all hypercollapsing subterms or inside one of them:

\begin{enumerate}[$\bullet$]

\item
If the redex occurs outside all hypercollapsing subterms, then $s_\delta \simhc s'_\delta$ together with orthogonality and fully-extendedness implies that a redex employing the same rewrite rule as the redex contracted in $s_\delta \rew t$ occurs at the same position in $s'_\delta$. Moreover, the redex occurs outside all hypercollapsing subterms by Proposition~\ref{prop:subreplaceokay}. Hence, contracting the redex in $s'_\delta$ yields a step $s'_\delta \outs t'$. That $t \simhc t'$ follows by $s_\delta \simhc s'_\delta$ and the fact that the same rewrite rule is employed in both $s_\delta \rew t$ and $s'_\delta \outs t'$: Clearly, $t$ and $t'$ are identical at all positions $p$ that descend from positions not in hypercollapsing subterms of $s_\delta$ or $s'_\delta$. If $q$ is the position of a maximal hypercollapsing subterm of $s_\delta$, then it is also the position of a maximal hypercollapsing subterm of $s'_\delta$ and vice versa, by Proposition \ref{prop:subreplaceokay}. The descendants of $q$ occur at identical positions in $t$ and $t'$ and are hypercollapsing subterms, since $s_\delta \simhc s'_\delta$ and since $\simhc$ is closed under substitution. Note, however, that the hypercollapsing subterms are not necessarily maximal.

\item
If the redex occurs inside a hypercollapsing subterm, then $t \simhc s_\delta$. Hence, by transitivity of $\simhc$ we have $t \simhc s'_\delta$ and we can define $t' = s'_\delta$.
\end{enumerate}

If $\gamma$ is a limit ordinal, the result is immediate by strong convergence and the induction hypothesis. \qed

Before proving the main theorem of this section, we show that reductions outside hypercollapsing subterms are confluent \emph{modulo} $\simhc$. To this end we first prove a restricted variant of the Strip Lemma. It is well-known that the usual Strip Lemma for iTRSs fails for \iLC \cite{KKSV97}, and, hence, we see that it must also fail for iCRSs.

\begin{lem}[Restricted Strip Lemma]
\label{lem:out_strip}
If $S: s \out t$ and $T: s \outs t'$, then $S/T$ and $T/S$ exist and end in the same term.
\end{lem}

\proof
Denote the length of $S$ by $\alpha$. We prove the lemma by ordinal induction on $\alpha$. Note that, since $T$ contracts a single redex $u$, we have that $T/S$ is actually a complete development of the residuals of $u$ in $t$. Obviously, if $\alpha = 0$, then the result follows trivially.

If $\alpha$ is a successor ordinal, then the result is immediate by Proposition \ref{prop:one_more} and the induction hypothesis.

If $\alpha$ is a limit ordinal, then Theorem \ref{the:tiling_completion_convergence} and the induction hypothesis ensure that we only need to show that T/S is strongly convergent. In other words, since $T$ contracts a single redex $u$, we need to prove that $u/S$ has a strongly convergent complete development. Assume the contrary and observe this implies the rewrite rule employed in $T$ is collapsing, otherwise any development of $u/S$ is strongly convergent.

By assumption, there exists a term $t^*$ in $T/S$ such that from $t^*$ onwards an infinite number of steps occur at a certain depth $d$ and no steps occur above $d$. Moreover, as function symbols have finite arity, there is a position $p$ at depth $d$ at which an infinite number of steps occur. As $T/S$ contracts only residuals of redexes in $t$, it follows by Lemma~\ref{lem:funky_bind} that redexes contracted along $T/S$ can only be nested by contracting a residual of a redex, say $v$, in $t$ such that $v$ occurs above all redexes in $t$ whose residuals are being nested. Hence, since only a finite number of residuals occurs in $t$ above the redex whose residual occurs at position $p$ in $t^*$, we have by the finite chains condition that the reducts of subterms of $t$ in the subterm at position $p$ in $t^*$ occur in finite chains. Hence, since again by Lemma~\ref{lem:funky_bind} no further nestings can be created among different reducts of the same subterm of $t$ or among reducts of parallel subterms of $t$, eventually all contracted redexes at $p$ are reducts of a single subterm in $t$. As there are an infinite number of steps at depth $d$, this means a hypercollapsing reduction exists starting in a subterm of $t$, say at position $q$.

By strong convergence and limit ordinal length of $S$, we can write $S = S_0 ; S_1$, where $S_0$ has successor ordinal length and $S_1 : s^* \out t$ is a non-empty final segment of $S$ contracting no redexes at prefix positions of $q$. Hence, $S_0$ has length strictly less than $\alpha$ and $s^*|_q \out t|_q$. As there is a hypercollapsing reduction starting from $t|_q$, it follows by Definition \ref{def:hyper_red} that there is also a hypercollapsing reduction starting from $s^*|_q$. But then, by Lemma \ref{lem:HCR->HC}, we have that $s^*|_q$ is hypercollapsing, which implies that $s^*|_q \out t|_q$ is empty and that $s^*|_q = t|_q$. Thus, $s^*|_q$ contains a set of descendants of $u$ having no complete development (giving rise to the hypercollapsing reduction starting from $s^*|_q = t|_q$), whence $u/S_0$ has no complete development. Since $S_0$ has length strictly less than $\alpha$, this contradicts the induction hypothesis. Hence, $T/S$ is strongly convergent. \qed

\begin{lem}
\label{lem:out_implies_closed}
If $s \out t_0$ and $s \out t_1$, then there exist terms $t^*_0$ and $t^*_1$ such that $t_0 \out t^*_0$ and $t_1 \out t^*_1$ with $t^*_0 \simhc t^*_1$.
\end{lem}

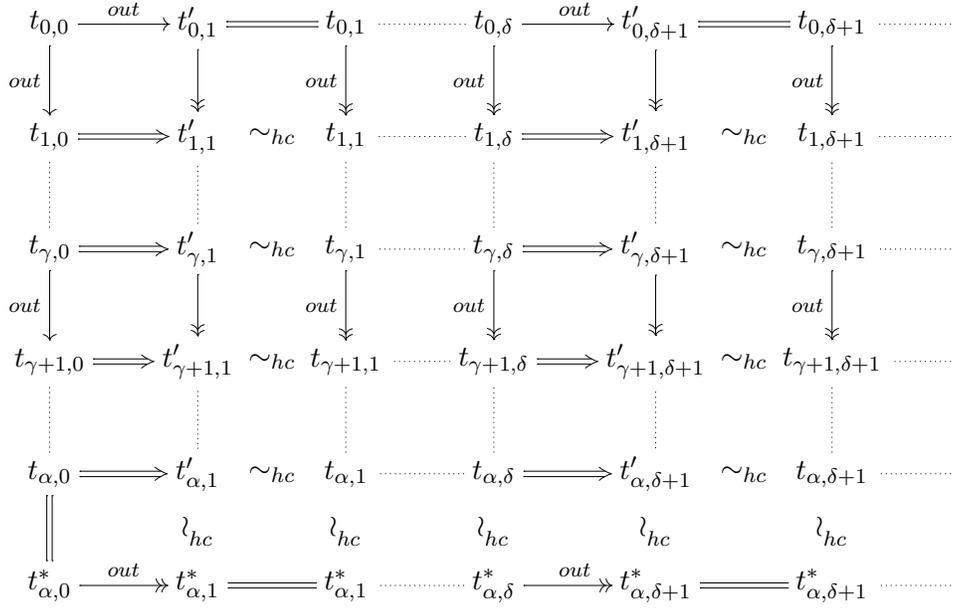
\begin{figure}
\centering
\subfloat[][]{
\xymatrix@=0.8cm{
s_{0, 0} \ar[r]^{out} \ar[d]_{out}
    & s_{0, 1} \ar@{.}[r] \ar@{=>}[d]
    & s_{0, \delta} \ar[r]^{out} \ar@{=>}[d]
    & s_{0, \delta + 1} \ar@{.}[r] \ar@{=>}[d]
    & s_{0, \beta} \ar@{=}[r] \ar@{=>}[d]
    & s^*_{0,\beta} \ar@{->>}[d]^{out} \\
s'_{1, 0} \ar@{->>}[r] \ar@{=}[d]
    & s'_{1, 1} \ar@{.}[r] \ar@{}[d]|*\txt{$\simhcv$}
    & s'_{1, \delta} \ar@{->>}[r] \ar@{}[d]|*\txt{$\simhcv$}
    & s'_{1, \delta + 1} \ar@{.}[r] \ar@{}[d]|*\txt{$\simhcv$}
    & s'_{1, \beta} \ar@{}[r]|*\txt{$\simhc$} \ar@{}[d]|*\txt{$\simhcv$}
    & s^*_{1,\beta} \ar@{=}[d]\\
s_{1, 0} \ar@{->>}[r]^{out} \ar@{.}[d]
    & s_{1, 1} \ar@{.}[r] \ar@{.}[d]
    & s_{1, \delta} \ar@{->>}[r]^{out} \ar@{.}[d]
    & s_{1, \delta + 1} \ar@{.}[r] \ar@{.}[d]
    & s_{1, \beta} \ar@{}[r]|*\txt{$\simhc$} \ar@{.}[d]
    & s^*_{1,\beta} \ar@{.}[d] \\
s_{\gamma, 0} \ar@{->>}[r]^{out} \ar[d]_{out}
    & s_{\gamma, 1} \ar@{.}[r] \ar@{=>}[d]
    & s_{\gamma, \delta} \ar@{->>}[r]^{out} \ar@{=>}[d]
    & s_{\gamma, \delta + 1} \ar@{.}[r] \ar@{=>}[d]
    & s_{\gamma, \beta} \ar@{}[r]|*\txt{$\simhc$} \ar@{=>}[d]
    & s^*_{\gamma,\beta} \ar@{->>}[d]^{out} \\
s'_{\gamma + 1, 0} \ar@{->>}[r] \ar@{=}[d]
    & s'_{\gamma + 1, 1} \ar@{.}[r] \ar@{}[d]|*\txt{$\simhcv$}
    & s'_{\gamma + 1, \delta} \ar@{->>}[r] \ar@{}[d]|*\txt{$\simhcv$}
    & s'_{\gamma + 1, \delta + 1} \ar@{.}[r] \ar@{}[d]|*\txt{$\simhcv$}
    & s'_{\gamma + 1, \beta} \ar@{}[r]|*\txt{$\simhc$}
                             \ar@{}[d]|*\txt{$\simhcv$}
    & s^*_{\gamma + 1,\beta} \ar@{=}[d] \\
s_{\gamma + 1, 0} \ar@{->>}[r]^{out} \ar@{.}[d]
    & s_{\gamma + 1, 1} \ar@{.}[r] \ar@{.}[d]
    & s_{\gamma + 1, \delta} \ar@{->>}[r]^{out} \ar@{.}[d]
    & s_{\gamma + 1, \delta + 1} \ar@{.}[r] \ar@{.}[d]
    & s_{\gamma + 1, \beta} \ar@{}[r]|*\txt{$\simhc$} \ar@{.}[d]
    & s^*_{\gamma + 1,\beta} \ar@{.}[d] \\
 & & & & &
}
\label{fig:hc_tile_a}
}
\newline
\subfloat[][]{
\xymatrix@=0.8cm{
t_{0, 0} \ar[d]_{out} \ar[r]^{out}
    & t'_{0, 1} \ar@{->>}[d] \ar@{=}[r]
    & t_{0, 1}  \ar@{->>}[d]_{out} \ar@{.}[r]
    & t_{0, \delta}  \ar@{->>}[d]_{out} \ar[r]^{out}
    & t'_{0, \delta + 1}  \ar@{->>}[d] \ar@{=}[r]
    & t_{0, \delta + 1}  \ar@{->>}[d]_{out} \ar@{.}[r]
    & \\
t_{1, 0} \ar@{.}[d] \ar@{=>}[r]
    & t'_{1, 1} \ar@{.}[d] \ar@{}[r]|*\txt{$\simhc$}
    & t_{1, 1} \ar@{.}[d] \ar@{.}[r]
    & t_{1, \delta} \ar@{.}[d] \ar@{=>}[r]
    & t'_{1, \delta + 1} \ar@{.}[d] \ar@{}[r]|*\txt{$\simhc$}
    & t_{1, \delta + 1} \ar@{.}[d] \ar@{.}[r]
    & \\
t_{\gamma, 0} \ar[d]_{out} \ar@{=>}[r]
    & t'_{\gamma, 1} \ar@{->>}[d] \ar@{}[r]|*\txt{$\simhc$}
    & t_{\gamma, 1} \ar@{->>}[d]_{out} \ar@{.}[r]
    & t_{\gamma, \delta} \ar@{->>}[d]_{out} \ar@{=>}[r]
    & t'_{\gamma, \delta + 1} \ar@{->>}[d] \ar@{}[r]|*\txt{$\simhc$}
    & t_{\gamma, \delta + 1} \ar@{->>}[d]_{out} \ar@{.}[r]
    & \\
t_{\gamma + 1, 0} \ar@{.}[d] \ar@{=>}[r]
    & t'_{\gamma + 1, 1} \ar@{.}[d] \ar@{}[r]|*\txt{$\simhc$}
    & t_{\gamma + 1, 1} \ar@{.}[d] \ar@{.}[r]
    & t_{\gamma + 1, \delta} \ar@{.}[d] \ar@{=>}[r]
    & t'_{\gamma + 1, \delta + 1} \ar@{.}[d] \ar@{}[r]|*\txt{$\simhc$}
    & t_{\gamma + 1, \delta + 1} \ar@{.}[d] \ar@{.}[r]
    & \\
t_{\alpha, 0} \ar@{=}[d] \ar@{=>}[r]
    & t'_{\alpha, 1} \ar@{}[d]|*\txt{$\simhcv$} \ar@{}[r]|*\txt{$\simhc$}
    & t_{\alpha, 1} \ar@{}[d]|*\txt{$\simhcv$} \ar@{.}[r]
    & t_{\alpha, \delta} \ar@{}[d]|*\txt{$\simhcv$} \ar@{=>}[r]
    & t'_{\alpha, \delta + 1} \ar@{}[d]|*\txt{$\simhcv$}
                              \ar@{}[r]|*\txt{$\simhc$}
    & t_{\alpha, \delta + 1} \ar@{}[d]|*\txt{$\simhcv$} \ar@{.}[r]
    & \\
t^*_{\alpha, 0} \ar@{->>}[r]^{out}
    & t^*_{\alpha, 1} \ar@{=}[r]
    & t^*_{\alpha, 1} \ar@{.}[r]
    & t^*_{\alpha, \delta} \ar@{->>}[r]^{out}
    & t^*_{\alpha, \delta + 1} \ar@{=}[r]
    & t^*_{\alpha, \delta + 1} \ar@{.}[r]
    &
}
\label{fig:hc_tile_b}
}
\caption{\label{fig:hc_tile}The `tiling diagrams' from the proof of Lemma \ref{lem:out_implies_closed}}
\end{figure}

\begin{figure}
\[
\xymatrix@!0@C=1.2cm@R=0.8cm{
*+[gray]{s_{\gamma, \delta}} \ar@{->>}@[gray][rrrr]^[gray]{out}
                                \ar@{=>}@[gray][dd]
                                \ar@{.}[dr]
    & & &
    & *+[gray]{s_{\gamma, \delta + 1}} \ar@{=>}@[gray]'[d][dd]
                                          \ar@{.}[dr] \\
& t_{\gamma, \delta} \ar@{=>}[rr] \ar@{->>}[dddd]_{out}
    &
    & t'_{\gamma, \delta + 1} \ar@{->>}[dddd]
    & \simhc & t_{\gamma, \delta + 1} \ar@{->>}[dddd]_{out} \\
*+[gray]{s'_{\gamma + 1, \delta}} \ar@{->>}@[gray]'[r]'[rrr][rrrr]
    & & &
    & *+[gray]{s'_{\gamma + 1, \delta + 1}} \\
*[gray]{\simhcv}
    & & &
    & *[gray]{\simhcv} \\
*+[gray]{s_{\gamma + 1, \delta}}
                            \ar@{->>}@[gray]'[r]'[rrr]^[gray]{out}[rrrr]
                            \ar@{.}[dr]
    & & &
    & *+[gray]{s_{\gamma + 1, \delta + 1}} \ar@{.}[dr] \\
& t_{\gamma + 1, \delta} \ar@{=>}[rr]
    &
    & t'_{\gamma + 1, \delta + 1} & \simhc & t_{\gamma + 1, \delta + 1}
}
\]
\caption{\label{fig:hc_overlay}Superimposing the `tiles' of the `tiling diagrams' in Figure \ref{fig:hc_tile}}
\end{figure}
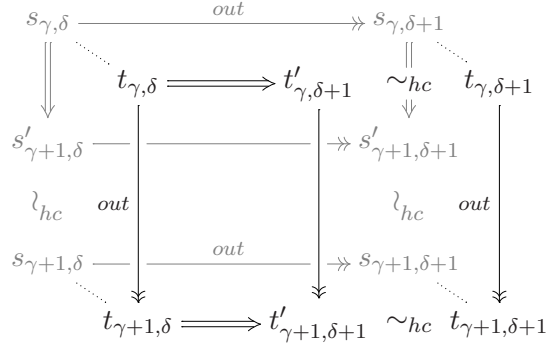

\proof
Let $S: s \out t_0$ and $T: s \out t_1$. By compression and Lemma \ref{hcoutlemma} we may assume that both $S$ and $T$ have length at most $\omega$. Suppose $S$ has length $\alpha \leq \omega$ and $T$ has length $\beta \leq \omega$. The proof proceeds in four steps: In the first step two `tiling diagrams' are constructed, yielding (i) a reduction starting in $t_0$, and (ii) a reduction starting in $t_1$. In the second step a relation is established between the `tiles' of the two diagrams. Employing the relation, it is shown in the third step that the two reductions obtained in the first step are strongly convergent. Finally, in the fourth step it is shown that the final terms of the two strongly convergent reductions are equivalent modulo $\simhc$. 

{\bf Tiling diagrams.}
Write $S : s_{0, 0} \outs s_{1, 0} \outs \cdots \, s_{\gamma, 0} \outs s_{\gamma + 1, 0} \outs \cdots \, s_{\alpha, 0}$ and $T : s_{0, 0} \outs s_{0, 1} \outs \cdots \, s_{0, \delta} \outs s_{0, \delta + 1} \outs \cdots \, s_{0, \beta}$ and define $s'_{\gamma, 0} = s_{\gamma, 0}$ for all $\gamma \leq \alpha$. We inductively construct the `tiling diagram' in Figure \ref{fig:hc_tile}\subref{fig:hc_tile_a}:
\begin{enumerate}[$\bullet$]
\item
the tiling of $s_{\gamma, 0} \outs s'_{\gamma + 1, 0}$ and $s_{\gamma, 0} \out s_{\gamma, \beta}$ exists by Lemma \ref{lem:out_strip};
\item
the reduction $s_{\gamma + 1, 0} \out s_{\gamma + 1, \beta}$ and the equivalences $s_{\gamma + 1, \delta} \simhc s'_{\gamma + 1, \delta}$ for all $0 \leq \delta \leq \beta$ exist by Lemma \ref{hcoutlemma} and the existence of $s'_{\gamma + 1, 0} \trewt s'_{\gamma + 1, \beta}$;
\item
the reduction $s^*_{\gamma, \beta} \out s^*_{\gamma + 1, \beta}$ and the equivalence $s'_{\gamma + 1, \beta} \simhc s^*_{\gamma + 1, \beta}$ exist by Lemma \ref{hcoutlemma} and the existence of $s_{\gamma, \beta} \trewt s'_{\gamma + 1, \beta}$;
\item
the equivalence $s^*_{\gamma + 1, \beta} \simhc s_{\gamma + 1, \beta}$ exists by transitivity of $\simhc$ and since $s_{\gamma + 1, \beta} \simhc s'_{\gamma + 1, \beta} \simhc s^*_{\gamma + 1, \beta}$. 
\end{enumerate}
As can be seen in Figure \ref{fig:hc_tile}\subref{fig:hc_tile_a}, the construction yields a reduction $S^*$ starting in $t_1 = s^*_{0, \beta}$ such that all steps in the reduction occur outside hypercollapsing subterms. Note that the constructed diagram is not a tiling diagram in the strict sense of the word: No reduction occurs at the bottom and the diagram consists not only of reductions but also of equivalences modulo hypercollapsing subterms.

To obtain the second `tiling diagram', depicted in Figure \ref{fig:hc_tile}\subref{fig:hc_tile_b}, we write $S : t_{0, 0} \outs t_{1, 0} \outs \cdots \, t_{\gamma, 0} \outs t_{\gamma + 1, 0} \outs \cdots \, t_{\alpha, 0}$ and $T : t_{0, 0} \outs t_{0, 1} \outs \cdots \, t_{0, \delta} \outs t_{0, \delta + 1} \outs \cdots \, t_{0, \beta}$ and define $t'_{0, \delta} = t_{0, \delta}$ for all $\delta \leq \beta$. The diagram is constructed by vertically repeating the horizontal construction of Figure  \ref{fig:hc_tile}\subref{fig:hc_tile_a}. The construction yields a reduction $T^* : t_0 = t^*_{\alpha, 0} \out t^*_{\alpha, 1} \out \cdots \, t^*_{\alpha, \delta} \out \cdots$.

{\bf Relation.}
Superimpose the tiles of the constructed `tiling diagrams' as depicted in Figure \ref{fig:hc_overlay}, i.e.\ $s_{\gamma, \delta}$ and $t_{\gamma', \delta'}$ are superimposed if $\gamma = \gamma'$ and $\delta = \delta'$. Define $s_{0, \delta} = s'_{0, \delta}$, and $t_{\gamma, 0} = t'_{\gamma, 0}$ for all $\gamma \leq \alpha$ and $\delta \leq \beta$. By construction of the `tiling diagrams', no term is superimposed on $s_{\gamma, \beta}$ with $\gamma \leq \alpha$ in case $\beta = \omega$ and similarly for $t_{\alpha, \delta}$ with $\delta < \beta$ in case $\alpha = \omega$.

We next prove for all superimposed terms $s_{\gamma, \delta}$ and $t_{\gamma, \delta}$ that $s_{\gamma, \delta} \simhc s'_{\gamma, \delta} \simhc t_{\gamma, \delta} \simhc t'_{\gamma, \delta}$. The proof is by induction on $\gamma$ and $\delta$. Induction is allowed because $s_{\gamma, \delta}$ and $t_{\gamma, \delta}$ exist for all $\gamma < \alpha$ and $\delta < \beta$:
\begin{enumerate}[$\bullet$]
\item
In case either $\gamma = 0$ \emph{or} $\delta = 0$, we have $s_{\gamma, \delta} = s'_{\gamma, \delta} = t_{\gamma, \delta} = t'_{\gamma, \delta}$ by definition. Hence, since $\simhc$ is an equivalence relation, $s_{\gamma, \delta} \simhc s'_{\gamma, \delta} \simhc t_{\gamma, \delta} \simhc t'_{\gamma, \delta}$.
\item
In case of $\gamma = \gamma' + 1$ \emph{and} $\delta = \delta' + 1$, we have by definition of the `tiling diagrams' that $s_{\gamma, \delta} \simhc s'_{\gamma, \delta}$ and $t_{\gamma, \delta} \simhc t'_{\gamma, \delta}$. Hence, by transitivity of $\simhc$, we obtain the desired result if we can establish $s_{\gamma, \delta} \simhc t'_{\gamma, \delta}$.

By Lemmas \ref{lem:out_strip} and \ref{hcoutlemma}, as employed in the construction of the `tiling diagrams', $s_{\gamma, \delta'} \out s_{\gamma, \delta}$ is essentially a development of residuals of the redex $u$ contracted in $s_{0, \delta'} \outs s_{0, \delta}$ such that no residuals of $u$ in $s_{\gamma, \delta}$ remain \emph{outside} hypercollapsing subterms. Since we have by the induction hypothesis that $s_{\gamma, \delta'} \simhc t_{\gamma, \delta'}$ and since every step in $s_{\gamma, \delta'} \out s_{\gamma, \delta}$ occurs outside hypercollapsing subterms, it follows by orthogonality and fully-extendedness that there exists a reduction $t_{\gamma, \delta'} \trewt t''_{\gamma, \delta}$ such that $s_{\gamma, \delta} \simhc t''_{\gamma, \delta}$. Since $s_{\gamma, \delta'} \out s_{\gamma, \delta}$ is essentially a development of residuals of $u$, it follows that $t_{\gamma, \delta'} \trewt t''_{\gamma, \delta}$ can be chosen to be a development of residuals of $u$, i.e.\ of the redex contracted in $t_{0, \delta'} \rew t'_{0, \delta}$. Moreover, it follows that all residuals of $u$ left in $t''_{\gamma, \delta}$ occur inside hypercollapsing subterms. Hence, since we have by Lemma~\ref{cdacrossd} that $t''_{\gamma, \delta} \trewt t'_{\gamma, \delta}$, we also have that $t''_{\gamma, \delta} \simhc t'_{\gamma, \delta}$. But then, by transitivity of $\simhc$ it follows that $s_{\gamma, \delta} \simhc t'_{\gamma, \delta}$, as required.
\end{enumerate}

{\bf Strong convergence.}
Employing that $s_{\gamma, \delta} \simhc s'_{\gamma, \delta} \simhc t_{\gamma, \delta} \simhc t'_{\gamma, \delta}$ holds for all superimposed $s_{\gamma, \delta}$ and $t_{\gamma, \delta}$, we next prove that the reduction $S^* : s^*_{0, \beta} \out s^*_{1, \beta} \out \cdots \, s^*_{\gamma, \beta} \out \cdots$ in Figure \ref{fig:hc_tile}\subref{fig:hc_tile_a} is strongly convergent. The proof is by contradiction. Thus, suppose $S^*$ is not strongly convergent. There now exists a position $p$ of minimal depth $d$ such that an infinite number of steps occur at $p$. As each step in $S^*$ occurs outside hypercollapsing subterms, it follows by minimality of $d$ that from some $\gamma$ onwards no redexes are contracted above $p$ and that all redexes contracted at $p$ are of non-collapsing rules. Moreover, by strong convergence of $s_{\gamma, 0} \trewt s_{\gamma, \beta}$, there is a $\delta$ such that all steps in $s_{\gamma, \delta} \trewt s_{\gamma, \beta}$ also occur below $d$.

Suppose for some minimal $\kappa \geq \gamma$ that a redex is contracted at some position $q < p$ in either $s_{\kappa, \delta} \dev s'_{\kappa + 1, \delta}$ or $s_{\kappa, \delta} \trewt s_{\kappa, \beta}$. By dependence of the depth of the steps in $s_{\kappa, \delta} \trewt s_{\kappa, \beta}$ on the depth of the steps in $s_{\lambda, \delta} \trewt s_{\lambda, \beta}$ for all $\gamma \leq \lambda < \kappa$, it follows by minimality of $\kappa$ that the reduction must be $s_{\kappa, \delta} \dev s'_{\kappa + 1, \delta}$. This implies that a redex is also contracted at position $q$ in $s_{\kappa, \beta} \dev s'_{\kappa + 1, \beta}$. Since the redex is by definition not contracted in $s^*_{\kappa, \beta} \out s^*_{\kappa + 1, \beta}$, it follows that the subterm at position $q$ in $s^*_{\kappa, \beta}$ is hypercollapsing. However, as $q < p$, this implies that the infinite number of redexes contracted at position $p$ cannot occur, as redexes in $S^*$ are contracted outside hypercollapsing subterms. Hence, for all $\kappa \geq \gamma$ we have that no reduction $s_{\kappa, \delta} \dev s'_{\kappa + 1, \delta}$ or $s_{\kappa, \delta} \trewt s_{\kappa, \beta}$ contracts a redex at strict prefix position of $p$.

Since all steps in $s_{\gamma, \delta} \trewt s_{\gamma, \beta}$ occur below $d$, the above implies that if a redex is contracted at position $p$ in some $s^*_{\kappa, \beta} \out s^*_{\kappa + 1, \beta}$ for minimal $\kappa \geq \gamma$, a redex is also contracted at position $p$ in $s_{\kappa, \delta} \dev s'_{\kappa + 1, \delta}$. Since the contracted redex is of a non-collapsing rule, it follows that the function symbol that occurs at position $p$ in both $s^*_{\kappa + 1, \beta}$ and $s'_{\kappa + 1, \delta}$ is the root symbol of the next redex contracted at position $p$. Hence, $s_{\gamma, \delta} \dev s'_{\gamma + 1, \delta} \simhc s_{\gamma + 1, \delta} \dev s'_{\gamma + 2, \delta} \simhc s_{\gamma + 2, \delta} \dev \cdots$ contains an infinite number of steps at position $p$ without any interleaving of collapsing steps at that position. However, as redexes contracted at position $p$ cannot occur inside hypercollapsing subterms by definition of $S^*$, we have that $t_{0, \delta} \trewt t_{\alpha, \delta}$ also contracts an infinite number of redexes at position $p$, which is impossible by strong convergence of this reduction, contradiction. Hence, $S^*$ is strongly convergent.

By a similar argument as above it follows that the reduction $T^* : t^*_{\alpha, 0} \out t^*_{\alpha, 1} \out \cdots \, t^*_{\alpha, \delta} \out \cdots$ is strongly convergent.

{\bf Equivalence modulo.}
Since $s_{\gamma, \delta} \simhc s'_{\gamma, \delta} \simhc t_{\gamma, \delta} \simhc t'_{\gamma, \delta}$ for all $\gamma$ and $\delta$ in both `tiling diagrams', the desired result follows by strong convergence. \qed

We can now --- finally --- prove the main result of the paper:
confluence modulo $\simhc$.

\begin{thm}
\label{the:confl_modulo}
Fully-extended, orthogonal iCRSs are confluent modulo $\simhc$.
\end{thm}

\proof
Let $s \simhc t$ and assume that $s \trewt s'$ and $t \trewt t'$. Consider the following diagram:
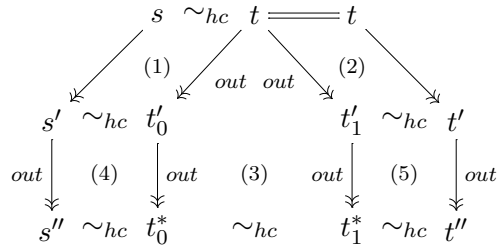
\[
\xymatrix{
    & s \ar@{}[r]|*\txt{$\simhc$} \ar@{->>}[dl] \ar@{}[d]|{\mathrm{(1)}}
    & t \ar@{=}[r] \ar@{->>}[dl]^{out} \ar@{->>}[dr]_{out}
    & t \ar@{->>}[dr] \ar@{}[d]|{\mathrm{(2)}} \\
s' \ar@{}[r]|*\txt{$\simhc$} \ar@{->>}[d]_{out} \ar@{}[dr]|{\mathrm{(4)}}
    & t'_0 \ar@{->>}[d]^{out} \ar@{}[drr]|{\mathrm{(3)}}
    & 
    & t'_1 \ar@{}[r]|*\txt{$\simhc$} \ar@{->>}[d]_{out}
    & t' \ar@{->>}[d]^{out} \ar@{}[dl]|{\mathrm{(5)}} \\
s'' \ar@{}[r]|*\txt{$\simhc$}
    & t^*_0 \ar@{}[rr]|*\txt{$\simhc$}
    &
    & t^*_1 \ar@{}[r]|*\txt{$\simhc$}
    & t''
}
\]
In the diagram, (1) and (2) exist by Lemma \ref{hcoutlemma} and (3) exists by Lemma \ref{lem:out_implies_closed}. Moreover, (4) and (5) also exist by Lemma \ref{hcoutlemma}. The result now follows by the diagram and transitivity of $\simhc$. \qed

\begin{exa}
The example iCRS from the introduction is confluent modulo $\simhc$
as it is orthogonal and fully-extended:

\begin{align*}
\mathtt{map}([z]F(z),\mathtt{cons}(X,XS)) & \rew
\mathtt{cons}(F(X),\mathtt{map}([z]F(z),XS)) \\
\mathtt{map}([z]F(z),\mathtt{nil}) & \rew \mathtt{nil}\\
\mathtt{hd}(\mathtt{cons}(X,XS)) & \rew X \\
\mathtt{tl}(\mathtt{cons}(X,XS)) & \rew XS
\end{align*}

The iCRS consisting of the first two rules above is confluent, because
it is confluent modulo~$\simhc$ and contains no collapsing rules.

The iCRS consisting of the infinite set of rules 
on the form 
\[
f^n([x]Z(x),Z') \rightarrow g(c^\omega,f^{n+1}([x]Z(x),Z(Z'))) \qquad n \geq 1
\]
is confluent modulo $\simhc$ (and confluent, as it does not contain collapsing rules).
\end{exa}

\subsection{Almost non-collapsingness}
\label{sec:almost_non_collapsing}

We would like to have a characterisation of confluence that appeals
only to the syntax of iCRSs without any need to consider 
equality modulo some relation. The first correct, fundamental
confluence result for iTRSs \cite{KKSV95} stated that 
an orthogonal iTRSs is confluent if{f} it has the property
of being `almost non-collapsing': There is at most one 
rule that is collapsing and the variable at the root of the
right-hand side of that rule is the only variable occurring in the left-hand
side of that rule.

Unfortunately, this concept does not carry over trivially to iCRSs,
when replacing the variables from iTRSs by meta-variables:
\begin{exa}
Consider the following rewrite rule, which is almost non-col\-laps\-ing in the above sense:
\[
f([x]Z(x)) \rew Z(f([x]Z(x)) \, .
\]
The term $f([x]f([y]x))$ gives rise to the finite reduction
\[
f([x]f([y]x)) \rew f([x]x) \, ,
\]
the final term of which reduces only to itself. However, the following reduction of length $\omega$ also exists:
\[
f([x]f([y]x)) \rew f([y]f([x]f([y]x))) \rew f([y]f([y]f([x]f([y]x)))) \rew \cdots \, s \, ,
\]
where $s$ is solution of the recursive equation $s = f([y]s)$, which is again a term which only reduces to itself. Hence, $f([x]f([y]x))$ reduces to two different terms that only reduce to themselves. In other words, the considered `almost non-collapsing' rewrite rule defines a non-confluent iCRS.
\end{exa}

We currently do not know how to give a precise characterisation of the class
of confluent iCRSs. From the above example, it is clear that almost
non-collapsingness alone does not suffice. It is plausible that the criterion
for confluence will be undecidable, even for the class of iCRSs containing
only a finite number of rules, all of which have finite right-hand-sides.
The above example bears witness of this: It crucially depends on the term
$f([x]x)$ being reachable from itself and reachability is of course in
general undecidable.

\section{Normal form properties}
\label{sec:normal_form_properties}

\noindent In this section we consider normal forms of iCRSs:
\begin{defi}
A term in an iCRS is a \emph{normal form} if no redexes occur in the term.
\end{defi}

The following properties relate normal forms and reductions. The properties extend their usual finitary counterparts to infinitary rewriting. Ample motivation for the formulation of the properties can be found in \cite{KKSV95}. In the definition, $\trewtc$ denotes the symmetric, transitive, reflexive closure of $\trewt$.

\begin{defi}
Define the following:
\begin{enumerate}[$\bullet$]
\item
An iCRS has the \emph{normal form property (NF)} if $s \trewtc t$ with $t$ a normal form implies $s \trewt t$.
\item
An iCRS has the \emph{unique normal form property (UN)} if $s \trewtc t$ with $s$ and $t$ normal forms implies $s = t$.
\item
An iCRS has the \emph{unique normal form property with respect to reduction (UN$^\rew$)} if $t \trewtb s \trewt t'$ with $t$ and $t'$ normal forms implies $t = t'$.
\end{enumerate}
\end{defi}

By the definitions we immediately have:
\begin{prop}
\label{prop:nornfun}
NF implies UN, and UN implies UN$^\rew$. \qed
\end{prop}

\begin{figure}
\centering
\subfloat[][]{
\xymatrix{
a \ar[r] \ar[d] & c \ar@(dr,dl) \\
b
}
\label{fig:nornfun_a}
}
\qquad \qquad
\subfloat[][]{
\xymatrix{
a_1 \ar[r] \ar[d] & c \ar@(dr,dl) & a_2 \ar[l] \ar[d] \\
b_1 & & b_2
}
\label{fig:nornfun_b}
}
\caption{\label{fig:nornfun}Counterexamples to the reverse of Proposition \ref{prop:nornfun}}
\end{figure}
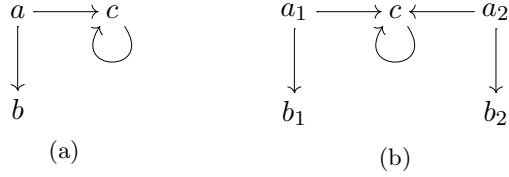

The converse implications of those above do not hold. This can be witnessed by the rewrite systems depicted in Figure \ref{fig:nornfun}.

In Figure \ref{fig:nornfun}\subref{fig:nornfun_a} we give a counterexample refuting that UN implies NF: As $b$ is the only normal form, next to all variables, UN is immediate. However, NF does not hold, as there is no reduction $c \trewt b$. The rewrite system in Figure \ref{fig:nornfun}\subref{fig:nornfun_b} refutes that UN$^\rew$ implies UN: Since $b_1$ is the only normal form of $a_1$ with respect to reduction and since $b_2$ the only normal form of $a_2$, UN$^\rew$ is immediate. However, UN does not hold, as we have $b_1 \trewtc b_2$, while $b_1 \not = b_2$.

The following lemma relates confluence modulo hypercollapsing subterms with the three properties introduced above.
\begin{lem}
\label{lem:UNR}
If an iCRS is confluent modulo hypercollapsing subterms, then NF, UN, and UN$^\rew$ hold.
\end{lem}

\proof
Let $s \trewtc t$ with $t$ a normal form. By induction on the number of changes in the direction of the reductions in $s \trewtc t$ and confluence modulo hypercollapsing subterms it follows that $s$ reduces to a term $t'$ such that $t \simhc t'$. As no hypercollapsing subterms occur in normal forms, we have $t = t'$. Hence, NF holds and UN and UN$^\rew$ follow by Proposition \ref{prop:nornfun}. \qed

In the above proof, confluence is easily substituted for confluence modulo hypercollapsing subterms, yielding the traditional result from finitary rewriting stating that confluence implies NF, UN, and UN$^\rew$. Moreover, as fully-extended, orthogonal iCRSs are confluent modulo hypercollapsing subterms by Theorem \ref{the:confl_modulo}, the above lemma also gives an affirmative answer to the conjecture posed in \cite{KV05} stating that fully-extended, orthogonal iCRSs satisfy UN$^\rew$.

It is not the case that NF implies confluence modulo hypercollapsing subterms. To see this, consider the following four rewrite rules:
\begin{align*}
a & \rew f(b) & b & \rew b\\
a & \rew g(c) & c & \rew c
\end{align*}
\noindent No term in which a redex occurs has a normal form. Hence, NF is immediate. However, confluence modulo hypercollapsing subterms does not hold, as $a$ reduces to $f(b)$ and $g(c)$ --- both of which only reduce to themselves --- and as $f(b) \not \simhc g(c)$.

\section{Conclusion and suggestions for future work}
\label{sec:conclusion}

\noindent We have extended confluence modulo identification of
hypercollapsing subterms to higher-order infinitary rewriting by
employing the proof techniques of earlier papers in the series on
iCRSs as well as extending the known proof methods from
\cite{T03_KV}. Our results properly generalise similar results for
iTRSs and \iLC, and the paper develops and extends the proof methods
employed in earlier papers on these subjects.

Two major open questions related to confluence of iCRSs and higher-order infinitary rewriting in general remain. We invite the reader to consider these: 

\begin{enumerate}[$\bullet$]

\item Can a characterisation be given of the subclass of confluent iCRSs, generalising the first-order result that almost-non-collapsing systems are confluent? As we reason in Section \ref{sec:almost_non_collapsing}, a generalisation will likely not be easy to come by.

\item The current proof of confluence modulo hypercollapsing subterms requires
orthogonality. Is it possible to replace orthogonality
by weak orthogonality?

\end{enumerate}

In the greater context of infinitary rewriting, this paper is part of an account of the general theory of iCRSs. We believe that the results and proof methods laid out will contribute to the further development of infinitary rewriting and equational reasoning involving infinite terms.

\section*{Acknowledgement}

\noindent The authors extend their thanks to the anonymous referees
for their diligent work and many comments that have led to substantial
improvements in the readability of the paper.

\bibliographystyle{abbrv}
\bibliography{icrs}

\begin{thebibliography}{10}

\bibitem{B85}
H.~P. Barendregt.
\newblock {\em The Lambda Calculus: Its Syntax and Semantics}.
\newblock Elsevier Science, revised edition, 1985.

\bibitem{C83}
B.~Courcelle.
\newblock Fundamental properties of infinite trees.
\newblock {\em Theoretical Computer Science}, 25(2):95--169, 1983.

\bibitem{GK94}
J.~Glauert and Z.~Khasidashvili.
\newblock Relative normalization in orthogonal expression reduction systems.
\newblock In {\em Proceedings of the International Workshop on Conditional (and
  Typed) Term Rewriting Systems (CTRS~'94)}, volume 968 of {\em Lecture Notes
  in Computer Science}, pages 144--165. Springer-Verlag, 1994.

\bibitem{HP96}
M.~Hanus and C.~Prehofer.
\newblock Higher-order narrowing with definitional trees.
\newblock In {\em Proceedings of the 7th International Conference on Rewriting
  Techniques and Applications (RTA~'96)}, volume 1103 of {\em Lecture Notes in
  Computer Science}, pages 138--152. Springer-Verlag, 1996.

\bibitem{K93}
S.~Kahrs.
\newblock Compilation of combinatory reduction systems.
\newblock In {\em Proceedings of the 1st International Workshop on Higher-Order
  Algebra, Logic, and Term Rewriting (HOA~'93)}, volume 816 of {\em Lecture
  Notes in Computer Science}, pages 169--188. Springer-Verlag, 1993.

\bibitem{KKSV97}
J.~R. Kennaway, J.~W. Klop, M.~R. Sleep, and F.-J. de~Vries.
\newblock Infinitary lambda calculus.
\newblock {\em Theoretical Computer Science}, 175(1):93--125, 1997.

\bibitem{T03_KV}
R.~Kennaway and F.-J. de~Vries.
\newblock Infinitary rewriting.
\newblock In Terese \cite{T03}, {C}hapter~12.

\bibitem{KKSV95}
R.~Kennaway, J.~W. Klop, R.~Sleep, and F.-J. de~Vries.
\newblock Transfinite reductions in orthogonal term rewriting systems.
\newblock {\em Information and Computation}, 119(1):18--38, 1995.

\bibitem{K08}
J.~Ketema.
\newblock On normalisation of infinitary combinatory reduction systems.
\newblock In {\em Proceedings of the 19th International Conference on Rewriting
  Techniques and Applications (RTA 2008)}, volume 5117 of {\em Lecture Notes in
  Computer Science}, pages 172--186. Springer-Verlag, 2008.

\bibitem{JJ05a}
J.~Ketema and J.~G. Simonsen.
\newblock Infinitary combinatory reduction systems.
\newblock In {\em Proceedings of the 16th International Conference on Rewriting
  Techniques and Applications (RTA 2005)}, volume 3467 of {\em Lecture Notes in
  Computer Science}, pages 438--452. Springer-Verlag, 2005.

\bibitem{JJ05b}
J.~Ketema and J.~G. Simonsen.
\newblock On confluence of infinitary combinatory reduction systems.
\newblock In {\em Proceedings of the 12th International Conference on Logic for
  Programming, Artificial Intelligence, and Reasoning (LPAR 2005)}, volume 3835
  of {\em Lecture Notes in Artificial Intelligence}, pages 199--214.
  Springer-Verlag, 2005.

\bibitem{paper_iii}
J.~Ketema and J.~G. Simonsen.
\newblock Infinitary combinatory reduction systems: Normalising reduction
  strategies, 2008.
\newblock Draft. Submitted for journal publication.

\bibitem{paper_i}
J.~Ketema and J.~G. Simonsen.
\newblock Infinitary combinatory reduction systems, 2009.
\newblock Draft. Submitted for journal publication.

\bibitem{K88}
Z.~Khasidashvili.
\newblock Beta-reductions and beta-developments of lambda-terms with the least
  number of steps.
\newblock In {\em Proceedings of the the International Conference in Computer
  Logic (COLOG~'88)}, volume 417 of {\em Lecture Notes in Computer Science},
  pages 105--111. Springer-Verlag, 1988.

\bibitem{K80}
J.~W. Klop.
\newblock {\em Combinatory Reduction Systems}.
\newblock PhD thesis, Rijksuniversiteit Utrecht, 1980.

\bibitem{KV05}
J.~W. Klop and R.~de~Vrijer.
\newblock Infinitary normalization.
\newblock In S.~N. Art{\"e}mov, H.~Barringer, A.~S. d'Avila Garcez, L.~C. Lamb,
  and J.~Woods, editors, {\em We Will Show Them: Essays in Honour of Dov
  Gabbay}, volume~2, pages 169--192. College Publications, 2005.

\bibitem{KOR93}
J.~W. Klop, V.~van Oostrom, and F.~van Raamsdonk.
\newblock Combinatory reduction systems: introduction and survey.
\newblock {\em Theoretical Computer Science}, 121(1 \& 2):279--308, 1993.

\bibitem{M97}
A.~Middeldorp.
\newblock Call by need computations to root-stable form.
\newblock In {\em Proceedings of the 24th Annual ACM SIGPLAN-SIGACT Symposium
  on Principles of Programming Languages (POPL~'97)}, pages 94--105, 1997.

\bibitem{SR90}
R.~C. Sekar and I.~V. Ramakrishnan.
\newblock Programming in equational logic: beyond strong sequentiality.
\newblock {\em Information and Computation}, 104(1):78--109, 1993.

\bibitem{T03}
Terese, editor.
\newblock {\em Term Rewriting Systems}, volume~55 of {\em Cambridge Tracts in
  Theoretical Computer Science}.
\newblock Cambridge University Press, 2003.

\bibitem{O96}
V.~van Oostrom.
\newblock Higher-order families.
\newblock In {\em Proceedings of the 7th International Conference on Rewriting
  Techniques and Applications (RTA~'96)}, volume 1103 of {\em Lecture Notes in
  Computer Science}, pages 392--407. Springer-Verlag, 1996.

\bibitem{O99}
V.~van Oostrom.
\newblock Normalisation in weakly orthogonal rewriting.
\newblock In {\em Proceedings of the 10th International Conference on Rewriting
  Techniques and Applications (RTA~'99)}, volume 1631 of {\em Lecture Notes in
  Computer Science}, pages 60--74. Springer-Verlag, 1999.

\bibitem{T03_R}
F.~van Raamsdonk.
\newblock Higher-order rewriting.
\newblock In Terese \cite{T03}, {C}hapter~11.

\end{thebibliography}

\newpage

\appendix

\section{Proof of Lemma \ref{lem:funky_bind}}
\label{app:bound}

We prove Lemma \ref{lem:funky_bind}.

\proof
Suppose that a variable bound by an abstraction in the redex pattern of $u_\alpha$ occurs in $v_\alpha$. We reason by transfinite induction on $\alpha$, the length of the reduction $s_0 \trewt^\alpha s_\alpha$. In case $\alpha = 0$, the result is immediate since bound variables can only occur below the abstraction by which they are bound and since subterms are substituted for the variables that occur in the subterm at the position of $u_0$ in $s_0$.

In case $\alpha$ is a successor ordinal, suppose either that (a) $p_0 \not < q_0$, or that (b) $q_\alpha \geq p_\alpha$, but $q_\alpha$ 
does not occur in the reduct $s_\alpha \vert_{p_\alpha}$ of $s_0 \vert_{p_0}$. We have the following:

\begin{enumerate}[$\bullet$]
\item
In case $p_0 \not < q_0$, we have that $v_0$ does not occur below $u_0$. Hence, for some $\beta < \alpha$ a nesting is created by contracting a redex at a prefix position of $p_\beta$ and $q_\beta$ and, by definition of valuations, if a variable is bound by the redex pattern of $u_{\beta + 1}$, then it cannot occur in $v_{\beta + 1}$.

\item
In case $q_\alpha \geq p_\alpha$, but $q_\alpha$ does not occur in the reduct of $s_0|_{p_0}$ at $p_\alpha$ in $s_\alpha$, there is a position $p_\alpha < p' \leq q_\alpha$ in $s_\alpha$ such that $s|_{p'}$ is a reduct of a subterm not strictly below $s_0 \vert_{p_0}$. Hence, for some $\beta < \alpha$ a nesting is created by contracting a redex at a prefix position of $p_\beta$ and $q_\beta$. In the term $s_{\beta + 1}$ a residual of $u_\beta$ occurs below $u_{\beta + 1}$ and above $v_{\beta + 1}$ and, by definition of valuations, if a variable is bound by the redex pattern of $u_{\beta + 1}$, then it cannot occur in the residual and, hence, in $v_{\beta + 1}$. 
\end{enumerate}

Thus, in both cases it follows for some $\beta < \gamma < \alpha$ that $s_\gamma \rew s_{\gamma + 1}$ nests a variable bound by $u_\gamma$ in $v_\gamma$. By the definition of valuations, we have for the redex contracted in $s_\gamma \rew s_{\gamma + 1}$, say $u'_\gamma$ at position $p'_\gamma$, that
\begin{enumerate}[(1)]
\item
a variable bound by an abstraction in the redex pattern of $u_\gamma$ occurs in $u'_\gamma$, and that
\item
a variable bound by an abstraction in the redex pattern of $u'_\gamma$ occurs in $v_\gamma$. 
\end{enumerate}
Since it follows by assumption that $u'_\gamma$ is the residual of a redex $u'_\delta$ in $s_\delta$ for all $\delta \leq \gamma$, we have by the induction hypothesis that $p'_\delta < q_\delta$ for all $\delta \leq \gamma$. Hence, for every $s_\delta \rew s_{\delta + 1}$ with $\delta \leq \gamma$ the contracted redex is a residual of a redex in $s_0$ in case it occurs at a prefix position of $p'_\gamma$. By the induction hypothesis we now have:
\begin{enumerate}[$\bullet$]
\item
In case $p_0 \not < q_0$, it follows that $p_0 < p'_0$ and $p'_0 < q_0$. Hence, $p_0 < q_0$, a contradiction.

\item
In case $q_\alpha \geq p_\alpha$, but $q_\alpha$ does not occur in the reduct of $s_0|_{p_0}$ at $p_\alpha$ in $s_\alpha$, it follows that $u'_{\beta + 1}$ occurs in the reduct of $s_0|_{p_0}$ at $p_{\beta + 1}$ in $s_{\beta + 1}$ and that $v_{\beta + 1}$ occurs in the reduct of $s_0|_{p'_0}$ at $p'_{\beta + 1}$ in $s_{\beta + 1}$. Hence, $v_{\beta + 1}$ occurs in the reduct of $s_0|_{p_0}$ at $p_{\beta + 1}$ in $s_{\beta + 1}$, a contradiction.
\end{enumerate}
Hence, the result follows if $\alpha$ is a successor ordinal.

In case $\alpha$ is a limit ordinal, the result is immediate by strong convergence and the induction hypothesis, since residuals occur at finite depth. \qed

\end{document}